\def\IEEEsubmission{0}

\def\complexNumbers{\mathbb{C}}
\def\realNumbers{\mathbb{R}}
\def\integers{\mathbb{Z}}

\def\constante{{\rm e}}
\def\constantj{{\rm j}}

\def\classicalMSEOperator[#1]{{{\text{MSE}}}\left(#1\right)}
\def\MSEOperator[#1]{{{\text{BMSE}}}\left(#1\right)}
\def\NMSEOperator[#1]{{\text{NBMSE}}\left(#1\right)}
\def\expectationOperator[#1][#2]{{\mathbb{E}_{#2}}\left[#1\right]}
\def\indicatorFunction[#1]{\mathbb{I}\left[{#1}\right]}
\def\probability[#1]{\textrm{Pr}\left({#1}\right)}
\def\complexGaussian[#1][#2]{\mathcal{CN}({#1,#2})}
\def\gaussianDist[#1][#2]{\mathcal{N}({#1,#2})}
\def\uniformDist[#1][#2]{\mathcal{U}({#1,#2})}
\def\binomDist[#1][#2]{\mathcal{B}({#1,#2})}
\def\gammaDist[#1][#2]{\Gamma({#1,#2})}
\def\varianceOperator[#1]{\textrm{var}\left(#1\right)}
\def\stdOperator[#1]{\textrm{std}\left[#1\right]}
\def\numberOfEdgeDevices{K}
\def\timeDomainOFDM[#1]{s(#1)}

\def\indexSubcarrier{l}
\def\numberOfActiveSubcarriers{M}

\def\dataSymbols[#1]{d_{#1}}

\def\receivedSymbolAtSubcarrier[#1]{\textbf{r}_{#1}^{(\indexCommunicationRound)}}
\def\transmittedSymbolAtSubcarrier[#1]{t_{#1}^{(\indexCommunicationRound)}}
\def\randomSymbolAtSubcarrier[#1]{s_{#1}^{(\indexCommunicationRound)}}
\def\channelAtSubcarrier[#1]{\textbf{h}_{#1}^{(\indexCommunicationRound)}}
\def\noiseAtSubcarrier[#1]{\textbf{n}_{#1}^{(\indexCommunicationRound)}}
\def\numberOfOFDMSymbols{S}
\def\indexOFDMSymbol{m}
\def\asymbolFromED[#1]{d_{#1}}

\def\exponentialIntegral[#1]{{\rm Ei}(#1)}
\def\tciFactor[#1]{p_{#1}}
\def\mappingFunction{\mathcal{M}}
\def\symbolEnergy{E_{\rm s}}
\def\indexActiveSymbol{\ell}
\def\voteInTime[#1][#2]{m_{#1,#2}}
\def\voteInFrequency[#1][#2]{l_{#1,#2}}
\def\numberOFEDsForOptionGeneral[\indexActiveSymbol]{\hat{U}^{+}_{\indexGradient,\indexDigit}}

\def\aFunction{f}
\def\vectorX{\textbf{\textrm{x}}}
\def\vectorY{\textbf{\textrm{y}}}

\def\numberOFEDsForOptionGeneral[#1]{K_{\indexGradient,\indexDigit,#1}}
\def\numberOFEDsForOptionGeneralDetector[#1]{\hat{K}_{\indexGradient,\indexDigit,#1}}
\def\numberOFEDsForOptionGeneralVar[#1]{{K}_{#1}}

\def\noiseVariance{\sigma_{\rm n}^2}

\def\coefficientOne{\sigma_{\rm channel}^2}
\def\coefficientTwo{\sigma_{\rm quan}^2}
\def\correctDecision[#1]{p_{#1}}
\def\incorrectDecision[#1]{q_{#1}}
\def\aparameterForBer[#1]{\epsilon_{#1}}

\def\probabilityIncorrect[#1]{P^{\rm err}_{#1}}

\def\oneVector[#1]{\textbf{\textrm{1}}_{#1}}
\def\zeroVector[#1]{\textbf{\textrm{0}}_{#1}}
\def\identityMatrix[#1]{{\textbf{\textrm{I}}_{#1}}}

\def\dataset[#1]{\mathcal{D}_{#1}}

\def\datasetBatch[#1]{\mathcal{\tilde{D}}_{#1}}
\def\batchSize{n_{\rm b}}

\def\completeData{\mathcal{D}}
\def\numberOfModelParameters{Q}
\def\sampleData[#1]{{\textrm{\textbf{x}}}_{#1}}
\def\sampleLabel[#1]{{y}_{#1}}
\def\learningRate{\eta}

\def\indexED{k}
\def\indexGradient{q}
\def\indexSampleData{{\ell}}
\def\indexCommunicationRound{\mathscr{t}}

\def\secondMoment[#1]{\delta_{#1}}

\def\modelParametersAtIteration[#1]{\textbf{w}^{(#1)}}
\def\modelParametersAtIterationEle[#1][#2]{w^{(#1)}_{#2}}

\def\modelParameters{\textbf{w}}
\def\modelParametersEle[#1]{{w}_{#1}}
\def\modelParametersOptimal{{\textbf{w}^{*}}}

\def\localGradientSign[#1][#2]{\bar{\textbf{g}}_{#1}^{(#2)}}
\def\localGradient[#1][#2]{\tilde{\textbf{g}}_{#1}^{(#2)}}
\def\localGradientNoIndex[#1]{\tilde{\textbf{g}}_{#1}}
\def\localGradientAll[#1][#2]{{\textbf{g}}_{#1}^{(#2)}}

\def\localGradientElementQuantized[#1][#2]{{\bar{\tilde{g}}}_{#1}^{(#2)}}
\def\localGradientElement[#1][#2]{{\tilde{g}}_{#1}^{(#2)}}
\def\localGradientNoIndexElement[#1]{{\tilde{g}}_{#1}}
\def\gradientWeight[#1]{\omega_{#1}}

\def\localGradientVectorQuantized[#1][#2]{{\bar{\tilde{\textbf{g}}}}^{(#2)}_{#1}}
\def\encoderGradient[#1][#2][#3]{\psi_{#1}^{#2}(#3)}

\def\lossFunctionSample[#1]{f(#1)}
\def\lossFunctionLocal[#1][#2]{F_{#1}(#2)}
\def\lossFunctionGlobal[#1]{F(#1)}
\def\gradientlossFunctionGlobal{\nabla F}

\def\lossFunctionGlobalMinimum{F^*}

\def\meanGradientEleEstimate[#1][#2]{{\hat{v}}^{(#1)}_{#2}}
\def\meanGradientEleOverQuantized[#1][#2]{{\bar{v}}^{(#1)}_{#2}}
\def\meanGradientEle[#1][#2]{{v}^{(#1)}_{#2}}
\def\meanGradientVector[#1]{\textbf{v}^{(#1)}}
\def\meanGradientVectorEstimate[#1]{\hat{\textbf{{v}}}^{(#1)}}
\def\meanGradientOverQuantized[#1]{{\bar{\textbf{v}}}^{(#1)}}

\def\globalGradient[#1]{{\textbf{{g}}}^{(#1)}}
\def\globalGradientElement[#1][#2]{{{g}}^{(#1)}_{#2}}

\def\globalGradientElementNoIndex[#1]{{g_{#1}}}
\def\bias[#1][#2]{{\textbf{{a}}}^{(#1)}_{#2}}
\def\biasQuan[#1]{{\textbf{{b}}}^{(#1)}}
\def\channelNoise[#1]{{\textbf{{c}}}^{(#1)}}

\def\configurationFactorChannel{E_{\rm channel}}
\def\configurationFactorQuan{E_{\rm quan}}
\def\configurationFactorTotal{E_{\rm total}}

\def\communicationRounds{T}

\def\metricForFirst[#1]{e_{#1}^{+}}
\def\metricForSecond[#1]{e_{#1}^{-}}

\def\LipschitzConstant{L}

\def\nonnegativeConstantsEle[#1]{L_{#1}}

\def\varianceBoundEle[#1]{\sigma_{#1}}

\def\channelVector[#1]{\textbf{\textrm{h}}_{#1}^{(\indexCommunicationRound)}}

\def\receiveVector[#1]{\textbf{\textrm{r}}_{#1}^{(\indexCommunicationRound)}}
\def\receiveVectorEstimate[#1]{\tilde{\textbf{\textrm{{r}}}}_{#1}^{(\indexCommunicationRound)}}

\def\matrixForCut[#1]{\textbf{\textrm{C}}_{#1}}
\def\symbolVector[#1]{\textbf{\textrm{d}}_{#1}^{(\indexCommunicationRound)}}
\def\symbolVectorEstimate[#1]{\tilde{\textbf{\textrm{{d}}}}_{#1}^{(\indexCommunicationRound)}}
\def\receivedVector[#1]{\textbf{\textrm{r}}_{#1}^{(\indexCommunicationRound)}}
\def\noiseVector[#1]{\textbf{\textrm{n}}_{#1}^{(\indexCommunicationRound)}}
\def\noiseVectorOnSymbols[#1]{\tilde{\textbf{\textrm{n}}}_{#1}^{(\indexCommunicationRound)}}

\def\transmittedVector[#1]{\textbf{\textrm{t}}_{#1}^{(\indexCommunicationRound)}}
\def\channelVector[#1]{\textbf{\textrm{h}}_{#1}^{(\indexCommunicationRound)}}
\def\idftMatrix[#1]{\textbf{\textrm{F}}_{#1}^{\rm H}}
\def\dftMatrix[#1]{\textbf{\textrm{F}}_{#1}}
\def\transformPrecoder[#1]{\textbf{\textrm{T}}_{#1}}
\def\transformDecoder[#1]{\textbf{\textrm{T}}_{#1}^{\rm H}}
\def\dftPrecoder[#1]{\textbf{\textrm{D}}_{#1}}
\def\dftDecoder[#1]{\textbf{\textrm{D}}_{#1}^{\rm H}}

\def\channelImpulseResponse[#1]{\textbf{\textrm{h}}_{#1}^{(\indexCommunicationRound)}}
\def\channelMatrix[#1]{\textbf{\textrm{H}}_{#1}^{(\indexCommunicationRound)}}
\def\channelMatrixDiag[#1]{{\bf \Lambda}_{#1}^{(\indexCommunicationRound)}}

\def\numberOfAntennasAtES{R}

\def\syncError{T_{\rm sync}}

\def\numberOfParametersPerOFDM{M_{\rm par}}

\def\distanceED[#1]{r_{#1}}
\def\powerED[#1]{P_{#1}}

\def\indexArea{u}
\def\Nerror{N_{\text{err}}}

\newcommand\mydots{\hbox to 1em{.\hss.\hss.}}

\if\IEEEsubmission1
\documentclass[journal,comsoc,12pt,onecolumn,draftclsnofoot,]{IEEEtran} 
\else
\documentclass[journal,comsoc]{IEEEtran}
\fi

\usepackage{multicol}
\usepackage{lipsum}
\usepackage{mathtools}
\usepackage{amsthm}
\usepackage{acronym}
\usepackage[bookmarksopen=true]{hyperref}
\usepackage{stfloats}
\usepackage{amsfonts}
\usepackage{acronym}
\usepackage{cite}
\usepackage{multirow}
\usepackage{bm}
\usepackage{amsmath,amssymb}
\usepackage{graphicx}
\usepackage[table,xcdraw]{xcolor}
\usepackage[caption=false,font=footnotesize]{subfig}
\usepackage[geometry]{ifsym}
\usepackage{array}
\usepackage[utf8]{inputenc}
\usepackage[T1]{fontenc}

\let\norm\undefined 
\DeclarePairedDelimiter\norm{\lVert}{\rVert}

\usepackage{tikz}
\usepackage{lipsum}
\usetikzlibrary{calc,shadings,patterns}
\makeatletter
\tikzset{%
  remember picture with id/.style={%
    remember picture,
    overlay,
    save picture id=#1,
  },
  save picture id/.code={%
    \edef\pgf@temp{#1}%
    \immediate\write\pgfutil@auxout{%
      \noexpand\savepointas{\pgf@temp}{\pgfpictureid}}%
  },
  if picture id/.code args={#1#2#3}{%
    \@ifundefined{save@pt@#1}{%
      \pgfkeysalso{#3}%
    }{
      \pgfkeysalso{#2}%
    }
  }
}

\def\savepointas#1#2{%
  \expandafter\gdef\csname save@pt@#1\endcsname{#2}%
}

\def\tmk@labeldef#1,#2\@nil{%
  \def\tmk@label{#1}%
  \def\tmk@def{#2}%
}

\tikzdeclarecoordinatesystem{pic}{%
  \pgfutil@in@,{#1}%
  \ifpgfutil@in@%
    \tmk@labeldef#1\@nil
  \else
    \tmk@labeldef#1,(0pt,0pt)\@nil
  \fi
  \@ifundefined{save@pt@\tmk@label}{%
    \tikz@scan@one@point\pgfutil@firstofone\tmk@def
  }{%
  \pgfsys@getposition{\csname save@pt@\tmk@label\endcsname}\save@orig@pic%
  \pgfsys@getposition{\pgfpictureid}\save@this@pic%
  \pgf@process{\pgfpointorigin\save@this@pic}%
  \pgf@xa=\pgf@x
  \pgf@ya=\pgf@y
  \pgf@process{\pgfpointorigin\save@orig@pic}%
  \advance\pgf@x by -\pgf@xa
  \advance\pgf@y by -\pgf@ya
  }%
}

\makeatother

\newcounter{hatchNumber}
\DeclarePairedDelimiter\floor{\lfloor}{\rfloor}

\usepackage{cuted}
\makeatletter
\newif\ifAC@uppercase@first%
\def\Aclp#1{\AC@uppercase@firsttrue\aclp{#1}\AC@uppercase@firstfalse}%
\def\AC@aclp#1{%
	\ifcsname fn@#1@PL\endcsname%
	\ifAC@uppercase@first%
	\expandafter\expandafter\expandafter\MakeUppercase\csname fn@#1@PL\endcsname%
	\else%
	\csname fn@#1@PL\endcsname%
	\fi%
	\else%
	\AC@acl{#1}s%
	\fi%
}%
\def\Acp#1{\AC@uppercase@firsttrue\acp{#1}\AC@uppercase@firstfalse}%
\def\AC@acp#1{%
	\ifcsname fn@#1@PL\endcsname%
	\ifAC@uppercase@first%
	\expandafter\expandafter\expandafter\MakeUppercase\csname fn@#1@PL\endcsname%
	\else%
	\csname fn@#1@PL\endcsname%
	\fi%
	\else%
	\AC@ac{#1}s%
	\fi%
}%
\def\Acfp#1{\AC@uppercase@firsttrue\acfp{#1}\AC@uppercase@firstfalse}%
\def\AC@acfp#1{%
	\ifcsname fn@#1@PL\endcsname%
	\ifAC@uppercase@first%
	\expandafter\expandafter\expandafter\MakeUppercase\csname fn@#1@PL\endcsname%
	\else%
	\csname fn@#1@PL\endcsname%
	\fi%
	\else%
	\AC@acf{#1}s%
	\fi%
}%
\def\Acsp#1{\AC@uppercase@firsttrue\acsp{#1}\AC@uppercase@firstfalse}%
\def\AC@acsp#1{%
	\ifcsname fn@#1@PL\endcsname%
	\ifAC@uppercase@first%
	\expandafter\expandafter\expandafter\MakeUppercase\csname fn@#1@PL\endcsname%
	\else%
	\csname fn@#1@PL\endcsname%
	\fi%
	\else%
	\AC@acs{#1}s%
	\fi%
}%
\edef\AC@uppercase@write{\string\ifAC@uppercase@first\string\expandafter\string\MakeUppercase\string\fi\space}%
\def\AC@acrodef#1[#2]#3{%
	\@bsphack%
	\protected@write\@auxout{}{%
		\string\newacro{#1}[#2]{\AC@uppercase@write #3}%
	}\@esphack%
}%
\def\Acl#1{\AC@uppercase@firsttrue\acl{#1}\AC@uppercase@firstfalse}
\def\Acf#1{\AC@uppercase@firsttrue\acf{#1}\AC@uppercase@firstfalse}
\def\Ac#1{\AC@uppercase@firsttrue\ac{#1}\AC@uppercase@firstfalse}
\def\Acs#1{\AC@uppercase@firsttrue\acs{#1}\AC@uppercase@firstfalse}

\newtheorem{theorem}{Theorem}
\newtheorem{definition}{Definition}
\newtheorem{lemma}{Lemma}

\newtheorem{assumption}{Assumption}
\newtheorem{remark}{\color{black} Remark} 
\newtheorem{example}{\color{black} Example} 
\DeclareMathOperator{\sign}{sign}
\def\signNormal[#1]{\sign\left(#1\right)}

\DeclareMathOperator{\diag}{diag}
\def\diagOperator[#1]{\diag\left\{#1\right\}}

\acrodef{QSGD}{quantized \ac{SGD}}
\acrodef{SNR}{signal-to-noise ratio}
\acrodef{RMSE}{root-mean-square error}
\acrodef{OFDM}{orthogonal frequency division multiplexing}
\acrodef{DFT}{discrete Fourier transform}
\acrodef{PSK}{phase-shift keying}
\acrodef{QAM}{quadrature amplitude modulation}
\acrodef{QPSK}{quadrature phase-shift keying}
\acrodef{PMEPR}{peak-to-mean envelope power ratio}
\acrodef{BER}{bit-error ratio}
\acrodef{SNR}{signal-to-noise ratio}
\acrodef{PSD}{power spectral density}
\acrodef{SE}{spectral efficiency}
\acrodef{CP}{cyclic prefix}
\acrodef{AWGN}{additive white Gaussian noise}
\acrodef{CFR}{channel frequency response}
\acrodef{CIR}{channel impulse response}
\acrodef{MMSE}{minimum mean square error}
\acrodef{LMMSE}{linear minimum mean square error}
\acrodef{BPSK}{binary phase shift keying}
\acrodef{BLER}{block-error rate}
\acrodef{ML}{maximum likelihood}
\acrodef{PHY}{physical layer}
\acrodef{PA}{power amplifier}
\acrodef{IDFT}{inverse DFT}
\acrodef{DoF}{degrees-of-freedom}
\acrodef{IoT}{Internet-of-Things}
\acrodef{FDE}{frequency-domain equalization}
\acrodef{FSK}{frequency-shift keying}
\acrodef{FSK-MV}{\ac{FSK}-based \ac{MV}}
\acrodef{PPM}{pulse-position modulation}
\acrodef{PPM-MV}{\ac{PPM}-based \ac{MV}}
\acrodef{RF}{radio-frequency}
\acrodef{IM}{index modulation}
\acrodef{BS}{base station}
\acrodef{MF}{matched filter}

\acrodef{BAA}{broadband analog aggregation}
\acrodef{OBDA}{one-bit broadband digital aggregation}
\acrodef{FEEL}{federated edge learning}
\acrodef{FL}{federated learning}
\acrodef{ED}{edge device}
\acrodef{ES}{edge server}
\acrodef{UL}{uplink}
\acrodef{DL}{downlink}
\acrodef{OAC}{over-the-air computation}
\acrodef{TCI}{truncated-channel inversion}
\acrodef{MV}{majority vote}
\acrodef{CNN}{convolution neural network}
\acrodef{ReLU}{rectified-linear unit}
\acrodef{CSI}{channel state information}
\acrodef{PAPR}{peak-to-average power ratio}
\acrodef{iid}[IID]{independent and identically distributed}
\acrodef{5G}{Fifth Generation}
\acrodef{4G}{Fourth Generation}
\acrodef{NR}{New Radio}
\acrodef{LTE}{Long Term Evolution}
\acrodef{RACH}{random-access channel}
\acrodef{DNN}{deep nueral network}
\acrodef{SGD}{stochastic gradient descent}
\acrodef{signSGD}{sign stochastic gradient descent}
\acrodef{5G}{Fifth Generation}
\acrodef{4G}{Fourth Generation}
\acrodef{NR}{New Radio}
\acrodef{LTE}{Long Term Evolution}
\acrodef{PRACH}{physical random access channel}
\acrodef{PUCCH}{physical uplink control channel}
\acrodef{OFDMA}{orthogonal frequency division multiple access}

\acrodef{MRC}{maximum-ratio combining}

\acrodef{CSK}{chirp-shift keying}
\acrodef{CSK-MV}[CSK-MV]{\ac{CSK}-based \ac{MV}}

\acrodef{MD}{multi-digit}
\acrodef{BMSE}{Bayesian \ac{MSE}}
\acrodef{MSE}{mean squared error}
\acrodef{NMSE}{normalized \ac{MSE}}
\acrodef{AAM}{adaptive absolute maximum}

\def\BibTeX{{\rm B\kern-.05em{\sc i\kern-.025em b}\kern-.08em
    T\kern-.1667em\lower.7ex\hbox{E}\kern-.125emX}}
\begin{document}

\title{Over-the-Air Computation Based on Balanced Number Systems for Federated Edge Learning
\\
\thanks{Alphan~\c{S}ahin is with the Electrical  Engineering Department,
	University of South Carolina, Columbia, SC, USA. E-mail: asahin@mailbox.sc.edu}
\author{Alphan~\c{S}ahin,~\IEEEmembership{Member,~IEEE}}
\thanks{{This paper in part was submitted to be presented  at the IEEE Global Communication Conference (GLOBECOM) 2022 - Workshop on Wireless Communications for Distributed Intelligence  \cite{sahin_2022gcWCDI}.}}
} 


\maketitle

\begin{abstract}
In this study, we propose a digital \ac{OAC} scheme for achieving continuous-valued (analog) aggregation for \ac{FEEL}.  We show that the average of a set of real-valued parameters can be calculated approximately by using the average of the corresponding numerals, where the numerals are obtained based on a balanced number system.  By exploiting this key property, the proposed scheme encodes the local stochastic gradients into a set of numerals. Next, it determines the positions of the activated  \ac{OFDM} subcarriers by using the values of the numerals. To eliminate the need for precise sample-level time synchronization, channel estimation overhead, and channel inversion, the proposed scheme also uses a non-coherent receiver at the \ac{ES} and does not utilize a pre-equalization at the \acp{ED}. We theoretically analyze the {MSE} performance of the proposed scheme and the convergence rate for a non-convex loss function. To improve the test accuracy of \ac{FEEL} with the proposed scheme, we  introduce the concept of \ac{AAM}.  Our numerical results show that  when the proposed scheme is used with \ac{AAM} for \ac{FEEL}, the test accuracy can reach up to 98\% for heterogeneous data distribution.
\end{abstract}
\acresetall
\section{Introduction}
\Ac{OAC} refers to the computation of mathematical functions by exploiting the signal-superposition property of wireless multiple-access channels \cite{Nazer_2007, Gastpar_2003}. To improve the utilization of limited wireless resources, it was initially considered for wireless sensor networks \cite{Goldenbaum_2013}.  With the same motivation, \ac{OAC} has recently gained increasing  attention in the literature for applications such as distributed learning  or wireless control systems \cite{Wanchun_2020,chen2021distributed,Park_2021}. For example, \ac{FEEL}, one of the promising distributed edge learning frameworks, aims to implement \ac{FL} \cite{pmlr-v54-mcmahan17a} over a wireless network. With \ac{FEEL}, the task of model training  is distributed across multiple \acp{ED} and the data uploading is avoided to promote user privacy \cite{Mingzhe_2021,chen2021distributed}. Instead of data samples, \acp{ED} share a large number of local stochastic gradients (or local model parameters) with an \ac{ES} for aggregation, e.g., averaging. However, typical orthogonal user multiplexing methods such as \ac{OFDMA} can be wasteful in this scenario since the \ac{ES} may not be interested in the local information of the \acp{ED} but only in a function of them. Similarly, a control system that requires an input that is a function of many \ac{IoT} devices' readings can suffer from high latency since the available spectrum for these networks is often limited and the \ac{OAC} can address the latency issues by calculating the functions, e.g., difference equations \cite{Park_2021}, over the air.


Although \ac{OAC} is a promising concept to address the latency issues in the aforementioned use cases, it is challenging to realize a reliable \ac{OAC} scheme under fading channels. 
This is because the typical equalization or phase correction methods used at the receiver for traditional multiple-access schemes, e.g., \ac{OFDMA}, cannot be directly employed for \ac{OAC}  to compensate for the channel distortion or imperfect synchronization as  the transmitted symbols are distorted by the channel  before the signal superposition. 
To address this issue, a majority of the state-of-the-art  \ac{OAC} methods rely on pre-equalization techniques\cite{Guangxu_2020, sery2020overtheair, Amiri_2020,Guangxu_2021, Zhang_2021, hellstrom2021overtheair,Liqun_2021,Zang_2020}. However, a pre-equalizer can impose stringent requirements on the underlying mechanisms such as time-frequency-phase synchronization, channel estimation, and channel prediction, which can be challenging to satisfy under non-stationary  channel conditions \cite{Careem_2020,Haejoon_2021}.
Most of the state-of-the-art \ac{OAC} schemes use analog modulation schemes to achieve continuous-valued computation, e.g.,  \cite{Guangxu_2020,Goldenbaum_2013tcom}, and \cite{Goldenbaum2010asilomar}. However,  in fading channels, analog modulations can be more susceptible to noise as compared to digital schemes. Although there are  digital methods, e.g., \ac{OBDA} \cite{Guangxu_2021} and \ac{FSK-MV} \cite{Sahin_2022MVjournal}, these schemes do not allow one to compute a continuous-valued function. In this paper, as opposed to earlier work, we introduce an \ac{OAC} scheme for \ac{FEEL}, where the local stochastic gradients are encoded into a set of numerals  based on a balanced (also called signed-digit) number system \cite{koren2018computer} to achieve a continuous-valued computation over  a digital scheme.\footnote{To avoid confusion, we use the terms "numeral" and "balanced" for "digit" and "signed-digit", respectively, since the term "digit" may specifically imply the ten symbols of the common base 10 numeral system.} The proposed method does not rely on pre-equalization and the availability of \ac{CSI} at the \acp{ED} and the \ac{ES}, leading to relaxed   synchronization requirements as compared to the methods relying on the availability of the \ac{CSI} at the \acp{ED}.

\subsection{Related Work}
\subsubsection{Over-the-air computation}
In the literature, \ac{OAC} schemes are particularly investigated to reduce the per-round communication latency of  \ac{FEEL}. 
 In \cite{Guangxu_2020}, \ac{BAA} that modulates  the \ac{OFDM} subcarriers with the model parameters is proposed. To overcome the impact of the multipath channel on the transmitted signals, the symbols on the \ac{OFDM} subcarriers are multiplied with the inverse of the channel coefficients and  the subcarriers that fade are excluded from the transmissions, known as {\em \ac{TCI}} in the literature. In \cite{sery2020overtheair}, an additional time-varying precoder is applied along with \ac{TCI} to facilitate the aggregation. 
In \cite{Amiri_2020}, the gradient estimates are sparsified and the sparse vectors are projected into a low-dimensional vector to reduce the bandwidth. The compressed data is transmitted with \ac{BAA}.
In \cite{hellstrom2021overtheair}, the power control and re-transmissions for BAA over static channels are investigated to obtain the optimal number of re-transmissions.
 In \cite{Liqun_2021}, instead of \ac{TCI}, the parameters are multiplied with the conjugate of the channel coefficients (i.e., maximum-ratio transmission) to increase the power efficiency. In \cite{Zang_2020}, the channel inversion is optimized with the consideration of sum-power constraint to avoid potential interference issues. In \cite{Guangxu_2021}, \ac{OBDA} is proposed  to facilitate the implementation of \ac{FEEL} for a practical wireless system. In this method,  considering distributed training by \ac{MV} with the  \ac{signSGD}~\cite{Bernstein_2018}, the \acp{ED} transmit \ac{QPSK} symbols  along with \ac{TCI}, where the real and imaginary parts of the \ac{QPSK} symbols are formed by using the signs of the stochastic gradients. 
At the \ac{ES}, the signs of the real and imaginary components of the superposed received symbols on each subcarrier are calculated to obtain the \ac{MV} for the sign of each gradient.
 The authors in \cite{Ruichen_2020} also consider \ac{OBDA}, but the pre-equalization in this method applies only phase correction to the transmitted symbols  (i.e., equal-gain transmission) by emphasizing the fact that amplitude alignment is not needed for digital \ac{OAC}.  
The reader is referred to \cite{Chen_2004} for various combining strategies for channel-aware decision fusion under the assumption of real-valued channel coefficients.

The methods relying on channel-inversion techniques require phase synchronization for coherent superposition. However, to achieve phase synchronization, a precise sample-level time synchronization at both EDs and ES needs to be maintained, which is very challenging in practice due to the  synchronization impairments and inaccurate clocks at the radios \cite{sahinDemo2022}. Also, residual carrier frequency causes additional phase rotations \cite{sahinSurvey2023}. To address these issues, in the literature, several \ac{OAC} methods that do not use pre-equalization are proposed at the expense of more resource consumption. For example, in \cite{weiICC_2022} and \cite{Amiria_2021}, the authors consider blind \acp{ED}, i.e., \ac{CSI} is not available at the \ac{ED}. By exploiting channel hardening, the \ac{ES} utilizes an estimate of the superposed \ac{CSI} to achieve an analog aggregation with \ac{MRC}. 
In \cite{Sahin_2022MVjournal, sahinCommnet_2021,sahinWCNC_2022}, and \cite{safiPIMRC_2022}, the \ac{OAC} for \ac{FEEL} is realized by exploiting non-coherent receiver techniques. Similar to \ac{OBDA} \cite{Guangxu_2021}, the schemes in these studies depend on the distributed training by  \ac{MV} \cite{Bernstein_2018}. However, instead of modulating the phase of the \ac{OFDM} subcarriers based on the sign of the local stochastic gradients, the schemes use  various keying approaches such as \ac{FSK}, \ac{PPM}, and \ac{CSK} along with a non-coherent comparator to detect the \ac{MV} at the \ac{ES}. While they can provide robustness against time variation of the wireless channel, synchronization errors, and imperfect power control, the 1-bit quantization nature of \ac{signSGD} can degrade the test accuracy in heterogeneous data distribution scenarios.  In \cite{Goldenbaum_2013tcom} and \cite{Goldenbaum2010asilomar}, Goldenbaum and Sta\'nczak  propose to calculate the energy of a sequence of superposed symbols.  In \cite{Goldenbaum_wcl2014}, they show that their scheme can also work when there is no \ac{CSI} at the transmitter under a scenario where the \ac{ES} is equipped with multiple
 antennas. 

\subsubsection{Quantization}

In the literature, extensive efforts have been made to decrease the communication costs of machine learning algorithms by quantization. For example, in \cite{Dan_2017},  a general \ac{QSGD} with the Elias integer encoding is investigated for encoding the  gradients by relying on the fact that large gradients are often less frequent.
The \ac{signSGD}, proposed in \cite{Bernstein_2018}, is an extreme case of quantization, where the signs of the gradients are considered for the training. In \cite{Jinjin_2020}, a ternary quantization, which is a balanced number system where the base is $3$, is applied to the model parameters to implement \ac{FL} based on parameter averaging. 
In \cite{kim2021tradeoff}, by considering the trade-off between precision and energy, the quantization levels for the neural
network parameters are optimized for \ac{FL}.
In \cite{liang2021wynerziv}, a gradient quantization method that uses the historical gradients as side information to compress the local gradients is proposed. The authors exploit the fact that 
gradients between adjacent rounds may have a high correlation for \ac{SGD}. 
Nevertheless, these quantization methods consider either ideal communication channels for training or orthogonal multiple access for \acp{ED}. Also, they do not consider an \ac{OAC} scheme.

\subsection{Contributions}
The contributions of this study can be listed as follows:

{\bf Continuous-valued \ac{OAC} with a digital scheme}: 
We show that the average of a set of real-valued parameters can be calculated by using the average of the corresponding numerals in the real domain, approximately. By exploiting this key property, discussed in Section~\ref{subsec:key}, we achieve continuous-valued computation over a digital \ac{OAC} scheme.  With the proposed method, the \acp{ED} first encode the real-valued local stochastic gradients into the numerals for a given balanced number system. The \acp{ED} activate the dedicated time-frequency resources (i.e., \ac{OFDM} subcarriers) based on the values of the numerals. The \acp{ED} simultaneously transmit their \ac{OFDM} symbols and the average numerals are calculated at the \ac{ES} with a non-coherent receiver. By using the average of the numerals, the \ac{ES} computes an estimate of the real-valued average stochastic gradient. To the best of our knowledge, this is the first study that uses a general balanced number system for \ac{OAC}. 

{\bf Theoretical MSE analysis:} We derive  the classical \ac{MSE} and \ac{BMSE}  of the estimator of the average stochastic gradient for a given set of parameters such as the number of numerals, number of \acp{ED},  and number of antennas at the \ac{ES}. By extending our initial work in \cite{sahin_2022gcWCDI}, we introduce the concept of \ac{AAM}, where each \ac{ED} shares a single parameter with the \ac{ES} to adjust the maximum quantization level to minimize the estimation error over the communication rounds of \ac{FEEL}.

{\bf Theoretical convergence analysis:} By using the \ac{MSE} derivation and considering both homogeneous and heterogeneous data distributions in the network, we show the convergence of \ac{FEEL} in the presence of the proposed scheme with and without \ac{AAM} for a non-convex loss function, i.e., Theorem~\ref{th:convergence} and Theorem~\ref{th:convergenceAAM}, respectively. While the proposed framework without \ac{AAM} contributes to the noise ball due to the stochastic gradients, the impact is largely addressed when the proposed scheme is utilized with the \ac{AAM}.

{\em Organization:} The rest of the paper is organized as follows. In Section \ref{sec:system}, the notation and the  preliminary discussions used in the rest of the sections are provided. 
In Section \ref{sec:dboac}, the proposed \ac{OAC} scheme and its \ac{MSE} performance are discussed. 
In Section~\ref{sec:conv}, the convergence rate of the \ac{FEEL} with the proposed scheme is discussed. In Section \ref{sec:numerical},  the numerical results are provided. We conclude the paper in Section \ref{sec:conclusion}.

{\em Notation:} The sets of complex numbers, real numbers, integers, and integers modulo $H$ are denoted by $\complexNumbers$,  $\realNumbers$, $\integers$, and $\integers_H$ respectively.
The $N$-dimensional all zero vector and the $N\times N$ identity matrix are  $\zeroVector[{N}]$ and $\identityMatrix[{N}]$, respectively. 
The function $\indicatorFunction[\cdot]$ results in $1$ if its argument holds, otherwise, it is $0$. 
$\expectationOperator[\cdot][x]$ and $\expectationOperator[\cdot][]$ are the expectation of its  argument over $x$ and the expectation of its  argument over all random variables, respectively. 
$ \nabla 
 \lossFunctionSample[{\modelParameters}]$ denotes the gradient of the function $f$, i.e. $\nabla f$, at the point $\modelParameters$. 
The zero-mean circularly symmetric multivariate  complex Gaussian distribution with the covariance matrix ${\textbf{\textrm{C}}_{\numberOfActiveSubcarriers}}$ of an $\numberOfActiveSubcarriers$-dimensional random vector $\vectorX\in\complexNumbers^{\numberOfActiveSubcarriers\times1}$ is denoted by
$\vectorX\sim\complexGaussian[\zeroVector[\numberOfActiveSubcarriers]][{\textbf{\textrm{C}}_{\numberOfActiveSubcarriers}}]$. The gamma distribution with the shape parameter $n$ and the rate  $\lambda$ is  $\gammaDist[n][\lambda]$.
The binomial distribution with the $K$ trials and the success probability  $p$ for each trial is $\binomDist[K][p]$. The uniform distribution with the support between $a$ and $b$ is $\uniformDist[a][b]$. Normal distribution with mean $\mu$ and variance $\sigma^2$ is $\gaussianDist[\mu][\sigma^2]$.
 The $\ell_2$-norm of the vector $\vectorX$ is $\norm{\vectorX}_2$. 

\section{Preliminaries and System Model}
\label{sec:system}
 In this section, we provide the signal and learning model that we use throughout the paper and the preliminaries related to encoding and decoding based on a balanced number system.
 
 \subsection{Signal Model}
We consider a wireless network with $\numberOfEdgeDevices$ \acp{ED} that are connected to an \ac{ES}, where each \ac{ED} and the \ac{ES}  are equipped with a single antenna and $\numberOfAntennasAtES$ antennas, respectively. We assume that the large-scale impact of the wireless channel is compensated with a power control mechanism, e.g., closed-loop power control with the \ac{PUCCH} in \ac{5G} \ac{NR} \cite{10.5555/3294673}, before the training process for \ac{FEEL} begins. 

For the signal model, we assume that the \acp{ED} access the wireless channel  on the same time-frequency resources {\em simultaneously} with $\numberOfOFDMSymbols$ \ac{OFDM} symbols consisting of $\numberOfActiveSubcarriers$ active subcarriers  for \ac{OAC}.  
 Assuming that the \ac{CP} duration is larger than the sum of the maximum time-synchronization error and the maximum-excess delay of the channel, we express the superposed modulation symbols on $\numberOfAntennasAtES$ antennas at the \ac{ES} for the $\indexSubcarrier$th subcarrier of the $\indexOFDMSymbol$th \ac{OFDM} symbol for the $\indexCommunicationRound$th communication  round of the training, i.e., $\receivedSymbolAtSubcarrier[{\indexSubcarrier,\indexOFDMSymbol}]\in\complexNumbers^{\numberOfAntennasAtES\times1}$ as
 \begin{align}
 	\receivedSymbolAtSubcarrier[{\indexOFDMSymbol,\indexSubcarrier}] = \sum_{\indexED=0}^{\numberOfEdgeDevices-1} \channelAtSubcarrier[\indexED,\indexOFDMSymbol,\indexSubcarrier]\transmittedSymbolAtSubcarrier[\indexED,\indexOFDMSymbol,\indexSubcarrier]+\noiseAtSubcarrier[\indexOFDMSymbol,\indexSubcarrier]~,
 	\label{eq:symbolOnSubcarrier}
 \end{align}
 where  $\channelAtSubcarrier[\indexED,\indexOFDMSymbol,\indexSubcarrier]\sim\complexGaussian[\zeroVector[\numberOfAntennasAtES]][{\identityMatrix[\numberOfAntennasAtES]}]$ is a $\numberOfAntennasAtES\times1$ vector that consists of the channel coefficients between $\numberOfAntennasAtES$ antennas at the \ac{ES} and the $\indexED$th \ac{ED}, $\transmittedSymbolAtSubcarrier[\indexED,\indexSubcarrier,\indexOFDMSymbol]\in\complexNumbers$ is the transmitted modulation symbol from the $\indexED$th \ac{ED}, and ${\noiseAtSubcarrier[\indexOFDMSymbol,\indexSubcarrier]}\sim\complexGaussian[\zeroVector[\numberOfAntennasAtES]][{\noiseVariance\identityMatrix[\numberOfAntennasAtES]}]$ is a $\numberOfAntennasAtES\times1$ \ac{AWGN} vector, where  $\noiseVariance$ is the noise variance for $\indexSubcarrier\in\integers_\numberOfActiveSubcarriers$ and $\indexOFDMSymbol\in\integers_\numberOfOFDMSymbols$.   We denote the \ac{SNR} of an \ac{ED} at the \ac{ES} receiver as $1/\noiseVariance$. 
 
 
  In practice, the synchronization point where the \ac{DFT} starts to be applied to the received signal for demodulation at the \ac{ES} and the time synchronization across the \acp{ED} may not be precise. To model former impairment, we assume that the synchronization point can deviate by $\Nerror$ samples within the \ac{CP} window. For the latter impairment, the time of arrivals of the \acp{ED}' signals at the \ac{ES} location are sampled from a uniform distribution between $0$ and $\syncError$~seconds, where $\syncError$ is equal to the reciprocal of the signal bandwidth. Note that the coarse time-synchronization can be maintained with the state-of-the-art protocols used in cellular systems. We introduce additional phase rotations to $\channelAtSubcarrier[\indexED,\indexOFDMSymbol,\indexSubcarrier]$ to capture the impact of the time-synchronization errors on $\receivedSymbolAtSubcarrier[{\indexOFDMSymbol,\indexSubcarrier}]$.  We assume that the frequency synchronization is handled before the transmissions with a control mechanism as done in 3GPP \ac{4G} \ac{LTE} and/or \ac{5G} \ac{NR} with \ac{RACH} and/or \ac{PUCCH}  \cite{10.5555/3294673} or custom methods such as AirShare \cite{Abari_2015}.

 \subsection{Learning Model}
 \label{subsec:feel}
 Let $\dataset[\indexED]$ denote the local data set containing the labeled data samples $(\sampleData[\indexSampleData], \sampleLabel[\indexSampleData] )$ at the $\indexED$th \ac{ED}, $\forall\indexED\in\integers_\numberOfEdgeDevices$,  where $\sampleData[\indexSampleData]$ is the $\indexSampleData$th data sample with its ground truth label $\sampleLabel[\indexSampleData]$. Suppose that all \acp{ED} upload their data sets to the \ac{ES}. The centralized learning problem can then  be expressed as 
 \begin{align}
 	\modelParametersOptimal=\arg\min_{\modelParameters\in\realNumbers^{\numberOfModelParameters}} \lossFunctionGlobal[\modelParameters]=\arg\min_{\modelParameters\in\realNumbers^{\numberOfModelParameters}} \frac{1}{|\completeData|}\sum_{\forall(\sampleData[\indexSampleData],\sampleLabel[\indexSampleData])\in\completeData} \lossFunctionSample[{\modelParameters;\sampleData[\indexSampleData],\sampleLabel[\indexSampleData]}]~,
 	\label{eq:clp}
 \end{align}
 where $\lossFunctionGlobal[\modelParameters]$ is the loss function,  $\completeData=\dataset[0]\cup\dataset[1]\cup\cdots\cup\dataset[\numberOfEdgeDevices-1]$ is the complete data set, and  $\lossFunctionSample[{\modelParameters;\sampleData[\indexSampleData],\sampleLabel[\indexSampleData]}]$ is the sample loss function for the parameters $\modelParameters=[\modelParametersEle[0],\mydots,\modelParametersEle[\numberOfModelParameters-1]]^{\rm T}\in\realNumbers^{\numberOfModelParameters}$, and $\numberOfModelParameters$ is the number of parameters. With (full-batch) gradient descent, a local optimum point can be  obtained as
 \begin{align}
 	\modelParametersAtIteration[\indexCommunicationRound+1] = \modelParametersAtIteration[\indexCommunicationRound] - \learningRate  \globalGradient[\indexCommunicationRound]~,
 	\label{eq:gd}
 \end{align}
 where $\learningRate$ is the learning rate and the gradient vector $\globalGradient[\indexCommunicationRound]=[\globalGradientElement[\indexCommunicationRound][0],\mydots,\globalGradientElement[\indexCommunicationRound][\numberOfModelParameters-1]]^{\rm T}\in\realNumbers^{\numberOfModelParameters}$ can be expressed as 
 \begin{align}
 	\globalGradient[\indexCommunicationRound] =  \nabla \lossFunctionGlobal[{\modelParametersAtIteration[\indexCommunicationRound]}]
 	= \frac{1}{|\completeData|}\sum_{\forall(\sampleData[\indexSampleData],\sampleLabel[\indexSampleData])\in\completeData} \nabla 
 	\lossFunctionSample[{\modelParametersAtIteration[\indexCommunicationRound];\sampleData[\indexSampleData],\sampleLabel[\indexSampleData]}]
 	~.
 	\label{eq:GlobalGradient}
 \end{align}
Equation \eqref{eq:gd} can be re-written as
\begin{align}
	\modelParametersAtIteration[\indexCommunicationRound+1] &= \modelParametersAtIteration[\indexCommunicationRound] - \learningRate \sum_{\indexED=0}^{\numberOfEdgeDevices-1}{\frac{|\dataset[\indexED]|}{|\completeData|}}\underbrace{\frac{1}{|\dataset[\indexED]|}\sum_{\forall(\sampleData[\indexSampleData],\sampleLabel[\indexSampleData]) \in\dataset[\indexED]} \nabla 
		\lossFunctionSample[{\modelParametersAtIteration[\indexCommunicationRound];\sampleData[\indexSampleData],\sampleLabel[\indexSampleData]}]}_{\triangleq\localGradientAll[\indexED][\indexCommunicationRound]}\nonumber\\&= \sum_{\indexED=0}^{\numberOfEdgeDevices-1}{\frac{|\dataset[\indexED]|}{|\completeData|}}\left(\modelParametersAtIteration[\indexCommunicationRound] - \learningRate \localGradientAll[\indexED][\indexCommunicationRound]\right) 	\nonumber
	~,
\end{align}
 where $\localGradientAll[\indexED][\indexCommunicationRound]\in\realNumbers^{\numberOfModelParameters}$ denotes the local gradient vector  at the $\indexED$th \ac{ED}. Therefore, \eqref{eq:gd} can still be realized by communicating the local gradients or locally updated model parameters between the \acp{ED} and the \ac{ES}, rather than moving the local data sets from the \acp{ED} to the \ac{ES}, which is beneficial for promoting data privacy \cite{Mingzhe_2021,chen2021distributed}. This observation also shows the underlying principle of the plain \ac{FL} based on gradient or model parameter aggregations \cite{pmlr-v54-mcmahan17a}. 
 
 \ac{FEEL} aims to realize \ac{FL} over a wireless network. In this study, we consider the implementation of \ac{FL} based on \ac{SGD}, known as FedSGD  \cite{pmlr-v54-mcmahan17a}, over a wireless network:  The $\indexED$th \ac{ED} calculates an estimate of the local gradient vector, denoted by $\localGradient[\indexED][\indexCommunicationRound]=[\localGradientElement[\indexED,0][\indexCommunicationRound],\mydots,\localGradientElement[\indexED,\numberOfModelParameters-1][\indexCommunicationRound]]^{\rm T}\in\realNumbers^{\numberOfModelParameters}$, as 
 \begin{align}
 	\localGradient[\indexED][\indexCommunicationRound] =  \nabla  \lossFunctionLocal[\indexED][{\modelParametersAtIteration[\indexCommunicationRound]}] 
 	= \frac{1}{\batchSize} \sum_{\forall(\sampleData[\indexSampleData],\sampleLabel[\indexSampleData])\in\datasetBatch[\indexED]} \nabla 
 	\lossFunctionSample[{\modelParametersAtIteration[\indexCommunicationRound];\sampleData[\indexSampleData],\sampleLabel[\indexSampleData]}]
 	~,
 	\label{eq:LocalGradientEstimate}
 \end{align}
 where $\datasetBatch[\indexED]\subset\dataset[\indexED]$ is the data batch obtained from the local data set and $\batchSize=|\datasetBatch[\indexED]|$ as the batch size. The \acp{ED} transmit the local gradient estimates to the \ac{ES}. Assuming identical data set sizes across the \acp{ED}, to solve \eqref{eq:clp}, the \ac{ES} calculates the average stochastic gradient vector $\meanGradientVector[\indexCommunicationRound]\triangleq[\meanGradientEle[\indexCommunicationRound][0],\mydots,\meanGradientEle[\indexCommunicationRound][\numberOfModelParameters-1]]^{\rm T}=\frac{1}{\numberOfEdgeDevices} \sum_{\indexED=0}^{\numberOfEdgeDevices-1}\localGradient[\indexED][\indexCommunicationRound]$ and broadcasts it to the \acp{ED}. Finally, the model parameters at the \acp{ED} are updated as
  \begin{align}
 	\modelParametersAtIteration[\indexCommunicationRound+1] = \modelParametersAtIteration[\indexCommunicationRound] - \learningRate\meanGradientVector[\indexCommunicationRound]~.
 	\label{eq:sgd}
 \end{align}

With traditional orthogonal user multiplexing, the per-round communication latency for \ac{FEEL} linearly increases with the number of \acp{ED} \cite{liu2021training}.
With the motivation of eliminating per-round communication latency, the main objective of this work is to calculate an estimate of $\meanGradientVector[\indexCommunicationRound]$, denoted by $\meanGradientVectorEstimate[\indexCommunicationRound]\triangleq[\meanGradientEleEstimate[\indexCommunicationRound][0],\mydots,\meanGradientEleEstimate[\indexCommunicationRound][\numberOfModelParameters-1]]^{\rm T}$, through a digital \ac{OAC} scheme robust against fading channels.

 \subsection{Balanced Number Systems}
\label{subsec:quanBalan}
\def\base{\beta}
\def\indexDigit{i}
\def\baseBT{\text{b3}}
\def\numberOfDigits{D}
\def\normalizationDigit{\xi}
\def\valueMaximum{v_{\rm max}}
\def\valueMaximumPrime{v_{\rm max}'}
\def\representationInBase[#1][#2]{\left(#2\right)}
\def\digitAveraged[#1][#2]{\mu_{{#1},{#2}}^{(\indexCommunicationRound)}}
\def\digitAveragedEst[#1][#2]{\hat{\mu}_{{#1},{#2}}^{(\indexCommunicationRound)}}
\def\digitGeneral[#1]{x_{#1}}
\def\digit[#1][#2]{d_{{#1},{#2}}^{(\indexCommunicationRound)}}
\def\digitStandart[#1]{b_{#1}}

\def\symbolSetWithoutZero{\mathbb{S}_\base^0}
\def\symbolWithoutZero[#1]{b_{#1}}
\def\resourceSet[#1]{\mathbb{T}_{#1}}

\def\symbolSet[#1]{\mathbb{S}_#1}
\def\symbol[#1]{a_{#1}}
\def\indexSymbol{j}
\def\decoder{f_{\text{dec},\base}}
\def\encoder{f_{\text{enc},\base}}
\def\aRealValue{v}
\def\aRealValueClamped{v'}
\def\aQuantizedValue{\bar{v}}
\def\encoderSB{f_{\text{enc}}^{\text{bin}}}
\def\encoderBT{f_{\text{enc}}^{\text{ter}}}
\def\stepSize{\Delta}
\def\functionBijection[#1]{f_{\rm bal}(#1)}
We define $\encoder$ as a function that maps $\aRealValue\in\realNumbers$ to a sequence  of $\numberOfDigits$ elements (i.e., $\numberOfDigits$ numerals in a balanced number system) in $\{\representationInBase[\base][{\digitGeneral[\numberOfDigits-1],\mydots,\digitGeneral[1],\digitGeneral[0]}]|\digitGeneral[\indexDigit]\in\symbolSet[\base],\base>1, \indexDigit\in\integers_\numberOfDigits\}$ as 
\begin{align}
	\representationInBase[\base][{\digitGeneral[\numberOfDigits-1],\mydots,\digitGeneral[1],\digitGeneral[0]}]=\encoder(\aRealValue)~,
\end{align}
where $\base$ is an odd positive integer (called base or scale \cite{hardy08}), $\digitGeneral[\indexDigit]$ is referred to as a numeral at the $\indexDigit$th position,  and $\symbolSet[\base]$ is the symbol set. Without loss of generality,
we define the symbol set $\symbolSet[\base]$ as 
\begin{align}
	\symbolSet[\base]
	\triangleq\{\symbol[{\indexSymbol}]|\symbol[{\indexSymbol}]=\functionBijection[\indexSymbol], \indexSymbol\in\integers_\base\}~,
	\label{eq:setDef}
\end{align} 
where $\functionBijection[\indexSymbol]$ is defined by
\begin{align}
	\functionBijection[\indexSymbol]\triangleq\begin{cases}
		-(\indexSymbol+1)/2, & \text{odd } \indexSymbol, \indexSymbol<\base-1\\
		(\indexSymbol+2)/2, & \text{even } \indexSymbol, \indexSymbol<\base-1\\
		0, & \indexSymbol=\base-1
	\end{cases}~.
	\label{eq:bal}
\end{align}
Based on \eqref{eq:bal}, $\symbol[{\base-1}]$ is a zero-valued symbol. The example symbol sets for $\base=5$ and $\base=7$ can obtained as $\symbolSet[5]=\{-1,1,-2,2,0\}$ and $\symbolSet[7]=\{-1,1,-2,2,-3,3,0\}$  , respectively.  For a balanced number system, there is no dedicated symbol for sign as $\symbolSet[\base]$ contains negative-valued symbols. 

The numerals are obtained via  $\encoder(\aRealValue)$  as follows: The encoder $\encoder(\aRealValue)$ first clamps $\aRealValue$
	for $\aRealValueClamped=\max(-\valueMaximum, \min(\aRealValue,\valueMaximum))$ to ensure $\aRealValueClamped\in[-\valueMaximum,\valueMaximum]$. It then re-scales $\aRealValueClamped$ as $\frac{\normalizationDigit}{\valueMaximum}\aRealValueClamped+\normalizationDigit+\frac{1}{2}$ and maps the scaled value to an integer between $0$ and $2\normalizationDigit$ with a floor operation  for   $\normalizationDigit\triangleq{(\base^{\numberOfDigits}-1)}/{2}$. Afterwards, it
	expands the corresponding integer as
	\begin{align}
		\floor*{\frac{\normalizationDigit}{\valueMaximum}\aRealValueClamped+\normalizationDigit+\frac{1}{2}}=\sum_{\indexDigit=0}^{\numberOfDigits-1}\digitStandart[\indexDigit]\base^\indexDigit~,
		\label{eq:unternaryConversion}
	\end{align}
	for $\digitStandart[\indexDigit]\in\integers_\base$, where  $(\digitStandart[\numberOfDigits-1],\mydots,\digitStandart[0])$ is the (unbalanced) base-$\base$ representation of the integer. Finally, it calculates  the numeral $\digitGeneral[\indexDigit]$ as $
	\digitGeneral[\indexDigit]=\digitStandart[\indexDigit]-(\base-1)/2\in\symbolSet[\base]$, $\forall\indexDigit$. 

\begin{example}
	\rm
	Assume that $\base=5$, $\numberOfDigits=3$, and $\valueMaximum=1$ and we want to calculate $\encoder(0.28)$ and $\encoder(-0.86)$. By the definition, $\normalizationDigit=(5^2-1)/2=62$.
	The base 5 representations of the decimal $\floor{62\times0.28+62+1/2}=79$ and the decimal $\floor{62\times-0.86+62+1/2}=9$  are $({\digitStandart[2]\digitStandart[1]\digitStandart[0]})_5=(304)_5$ and $({\digitStandart[2]\digitStandart[1]\digitStandart[0]})_5=(014)_5$, respectively. Since $
	\digitGeneral[\indexDigit]\triangleq\digitStandart[\indexDigit]-(\base-1)/2$, we obtain $\encoder(0.28)=\representationInBase[5][1,-2,2]$, and $\encoder(-0.86)=\representationInBase[5][-2,-1,2]$.
	\label{ex:enc}
\end{example}

The corresponding decoder $\decoder$ that maps the sequence $\representationInBase[\base][{\digitGeneral[\numberOfDigits-1],\mydots,\digitGeneral[1],\digitGeneral[0]}]$  to $\aQuantizedValue\in\realNumbers$ can be expressed as
\begin{align}
	\aQuantizedValue=
	\decoder\representationInBase[\base][{\digitGeneral[\numberOfDigits-1],\mydots,\digitGeneral[1],\digitGeneral[0]}]\triangleq\frac{\valueMaximum}{\normalizationDigit}\sum_{\indexDigit=0}^{\numberOfDigits-1}\digitGeneral[\indexDigit]\base^{\indexDigit}~.
	\label{eq:decoderFcn}
\end{align}
\begin{example}
	\rm
	Consider the parameters given in Example~\ref{ex:enc}. Hence, we obtain $\decoder\encoder(0.28)=\decoder\representationInBase[5][1,-2,2]\approxeq0.2742$, and $\decoder\encoder(-0.86)=\decoder\representationInBase[5][-2,-1,2]\approxeq-0.8548$ based on \eqref{eq:decoderFcn}. The step size can also be calculated as $\stepSize=2/(5^3-1)\approxeq0.016$. 
	\label{ex:dec}
\end{example}
Note that $\aQuantizedValue=\decoder\encoder(\aRealValue)$ forms a mid-tread uniform quantization, i.e., zero is one of the re-construction levels. The quantization step size can also be calculated as $\stepSize=2\valueMaximum/(\base^\numberOfDigits-1)$ and the quantization error, i.e., $|{\aRealValue-\aQuantizedValue}|$, decreases with increasing $\numberOfDigits$ for $|\aRealValue|\le\valueMaximumPrime$ for $\valueMaximumPrime=\valueMaximum+\stepSize/2$. 

\begin{figure}[t]
	\centering
	{\includegraphics[width =3.3in]{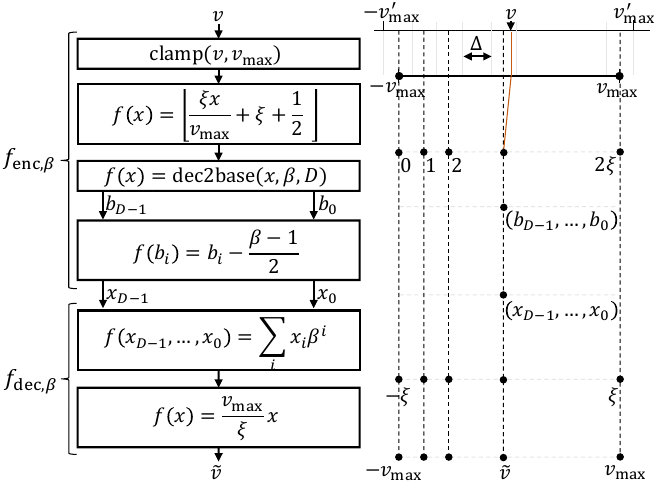}
	} 
	\caption{The operations in $\encoder$ and $\decoder$ and the corresponding input-output relationships.}
	\label{fig:encoderDecoder}
\end{figure}

The operations in $\encoder$ and $\decoder$ and the corresponding input-output relationships are given for an arbitrary input in \figurename~\ref{fig:encoderDecoder}. We utilize $\encoder$ and  $\decoder$  to encode the local stochastic gradients at the EDs and obtain an estimate of the arithmetic mean of the local stochastic gradients at the ES, respectively,  as discussed in Section~\ref{sec:dboac}.

\section{Proposed OAC Scheme}
\label{sec:dboac}
In this section, we discuss the proposed \ac{OAC} scheme relying on the representation of the gradients based on a balanced number system.  We analyze its performance in terms of \ac{MSE} and introduce the \ac{AAM} to improve the \ac{MSE} over the communication rounds of \ac{FEEL}.

\subsection{Key Observation}
\label{subsec:key}
Based on the discussions given in Section~\ref{subsec:feel}, consider the $\indexGradient$th gradient at the $\indexED$th \ac{ED} for the $\indexCommunicationRound$th communication round of the \ac{FEEL}, i.e., $\localGradientElement[\indexED,\indexGradient][\indexCommunicationRound]$. Suppose that  $\localGradientElement[\indexED,\indexGradient][\indexCommunicationRound]$ is encoded into the sequence of length $\numberOfDigits$ denoted by
\begin{align}
\representationInBase[\base][{\digit[\indexED,\indexGradient][\numberOfDigits-1],\mydots,\digit[\indexED,\indexGradient][1],\digit[\indexED,\indexGradient][0]}]=\encoder(\localGradientElement[\indexED,\indexGradient][\indexCommunicationRound])~,
\label{eq:encodedSequence}
\end{align}
for $\digit[\indexED,\indexGradient][\indexDigit]\in\symbolSet[\base]$.
By using the definition of $\decoder$ in \eqref{eq:decoderFcn}, the $\indexGradient$th average stochastic gradient, i.e., $\meanGradientEle[\indexCommunicationRound][\indexGradient]=
\frac{1}{\numberOfEdgeDevices}\sum_{\indexED=0}^{\numberOfEdgeDevices-1} \localGradientElement[\indexED,\indexGradient][\indexCommunicationRound]$,  can be obtained approximately as
\begin{align}
	\meanGradientEle[\indexCommunicationRound][\indexGradient]\approxeq\meanGradientEleOverQuantized[\indexCommunicationRound][\indexGradient]\triangleq\frac{1}{\numberOfEdgeDevices}\sum_{\indexED=0}^{\numberOfEdgeDevices-1}& \localGradientElementQuantized[\indexED,\indexGradient][\indexCommunicationRound]=\frac{\valueMaximum}{\normalizationDigit}\sum_{\indexDigit=0}^{\numberOfDigits-1} \underbrace{\frac{1}{\numberOfEdgeDevices}\sum_{\indexED=0}^{\numberOfEdgeDevices-1}\digit[\indexED,\indexGradient][\indexDigit]}_{\triangleq\digitAveraged[\indexGradient][\indexDigit]}\base^{\indexDigit}\nonumber
	\\
	&=\decoder\representationInBase[\base][{\digitAveraged[\indexGradient][\numberOfDigits-1],\mydots,\digitAveraged[\indexGradient][1],\digitAveraged[\indexGradient][0]}]~,
	\label{eq:basicAveraging}
\end{align}
where $\localGradientElementQuantized[\indexED,\indexGradient][\indexCommunicationRound]$ is the quantized gradient, i.e.,
$\localGradientElementQuantized[\indexED,\indexGradient][\indexCommunicationRound] = \decoder\encoder(\localGradientElement[\indexED,\indexGradient][\indexCommunicationRound])$.

Equation \eqref{eq:basicAveraging} implies that $\meanGradientEle[\indexCommunicationRound][\indexGradient]$ can be calculated approximately by evaluating the function $\decoder$ with the values that are calculated by averaging the numerals across $\numberOfEdgeDevices$ \acp{ED} in the {\em real} field, i.e., $\{\digitAveraged[\indexGradient][\indexDigit]|\indexDigit\in\integers_\numberOfDigits\}$. By evaluating $\digitAveraged[\indexGradient][\indexDigit]$ further, it can also be shown that
\begin{align}
\digitAveraged[\indexGradient][\indexDigit] = \frac{1}{\numberOfEdgeDevices}\sum_{\indexED=0}^{\numberOfEdgeDevices-1}\digit[\indexED,\indexGradient][\indexDigit] = \frac{1}{\numberOfEdgeDevices}\sum_{\indexSymbol=0}^{\base-1} \symbol[\indexSymbol]\numberOFEDsForOptionGeneral[\indexSymbol]~,
\label{eq:digitAveraging}
\end{align}
where $\numberOFEDsForOptionGeneral[\indexSymbol]$ denotes the number of \acp{ED} with the symbol $\symbol[\indexSymbol]$ for the $\indexDigit$th numeral in \eqref{eq:encodedSequence} and the $\indexGradient$th gradient. Note that the identity in \eqref{eq:digitAveraging} is due to the definition of expectation for discrete outcomes as given for a probability mass function.

\begin{example}
	\rm 
Assume that $\numberOfEdgeDevices=2$, $\localGradientElement[0,\indexGradient][\indexCommunicationRound]=0.28$, and $\localGradientElement[1,\indexGradient][\indexCommunicationRound]=-0.86$. The average of the gradients can be calculated as $\meanGradientEle[\indexCommunicationRound][\indexGradient]=(\localGradientElement[0,\indexGradient][\indexCommunicationRound]+\localGradientElement[1,\indexGradient][\indexCommunicationRound])/2=-0.29$. Now, consider the encoder parameters given in Example~\ref{ex:enc}. We obtain $\encoder(0.28)=\representationInBase[5][1,-2,2]$, and $\encoder(-0.86)=\representationInBase[5][-2,-1,2]$. Therefore, the average of the numerals can be calculated as $(\digitAveraged[\indexGradient][2],\digitAveraged[\indexGradient][1],\digitAveraged[\indexGradient][0])=(1-2,-2-1,2+2)/2=(-1/2,-3/2,2)$. Also, notice that $(\digitAveraged[\indexGradient][2],\digitAveraged[\indexGradient][1],\digitAveraged[\indexGradient][0])$ can be calculated by using the number of \acp{ED} that votes for each element of $\{-1,1,-2,2,0\}$. For instance, $\digitAveraged[\indexGradient][0]$ can be calculated via the last expression in \eqref{eq:digitAveraging} for
$(\numberOFEDsForOptionGeneral[0],\numberOFEDsForOptionGeneral[1],\numberOFEDsForOptionGeneral[2],\numberOFEDsForOptionGeneral[3],\numberOFEDsForOptionGeneral[4])=(0,0,0,2,0)$ where the corresponding symbols are $(\symbol[0],\symbol[1],\symbol[2],\symbol[3],\symbol[4])=(-1,1,-2,2,0)$ for $\base=5$.
By evaluating $\meanGradientEleOverQuantized[\indexCommunicationRound][\indexGradient]=\decoder(-1/2,-3/2,2)$, we obtain $\meanGradientEleOverQuantized[\indexCommunicationRound][\indexGradient]\approxeq-0.2903$. Note $\meanGradientEleOverQuantized[\indexCommunicationRound][\indexGradient]$ is also equal to the average of the quantized gradients, i.e., $\decoder\encoder(0.28)\approxeq0.2742$ and $\decoder\encoder(-0.86)\approxeq-0.8548$, as exemplified in Example~\ref{ex:dec}.  
\label{ex:sum}
\end{example}

The proposed \ac{OAC} scheme  computes an estimate of $\meanGradientEle[\indexCommunicationRound][\indexGradient]$ by relying on the expansion in \eqref{eq:basicAveraging} and the identity given in \eqref{eq:digitAveraging}, rather than averaging the continuous $\localGradientElement[\indexED,\indexGradient][\indexCommunicationRound]$ with an analog \ac{OAC} such as \ac{BAA} proposed in \cite{Guangxu_2020} or Goldenbaum's scheme in \cite{Goldenbaum_2013tcom}.

\subsection{Edge Device - Transmitter}
\label{subsec:tx}
At the $\indexCommunicationRound$th communication round of the \ac{FEEL}, the $\indexED$th \ac{ED} calculates the numerals $\{\digit[\indexED,\indexGradient][\indexDigit]|\indexGradient\in\integers_\numberOfModelParameters, \indexDigit\in\integers_\numberOfDigits\}$ with \eqref{eq:encodedSequence}, for a given $\base$. 
The main strategy exploited at  the $\indexED$th \ac{ED} with the proposed scheme is that {\em$\base-1$ subcarriers are dedicated for each numeral  and one of them is  activated  based on its value}.
To express this encoding operation rigorously, let $\mappingFunction$ be a function that maps $\indexGradient\in\integers_\numberOfModelParameters$ to a set of $(\base-1)\numberOfDigits$ distinct time-frequency index pairs denoted by $\resourceSet[\indexGradient]\triangleq\{(\voteInTime[\indexDigit][\indexActiveSymbol],\voteInFrequency[\indexDigit][\indexActiveSymbol])|\indexDigit\in\integers_\numberOfDigits,\indexActiveSymbol\in\integers_{\base-1}\}$ for $\voteInTime[\indexDigit][\indexActiveSymbol]\in\integers_\numberOfOFDMSymbols$ and $\voteInFrequency[\indexDigit][\indexActiveSymbol]\in\integers_\numberOfActiveSubcarriers$, where $\resourceSet[\indexGradient_1]\cap\resourceSet[\indexGradient_2]=\emptyset$ if $\indexGradient_1\neq\indexGradient_2$ for $\indexGradient_1,\indexGradient_2\in\integers_\numberOfModelParameters$.
 The $\indexED$th \ac{ED} determines the modulation symbol $\transmittedSymbolAtSubcarrier[\indexED,{\voteInTime[\indexDigit][\indexActiveSymbol],\voteInFrequency[\indexDigit][\indexActiveSymbol]}]$  as
\begin{align}
	\transmittedSymbolAtSubcarrier[\indexED,{\voteInTime[\indexDigit][\indexActiveSymbol],\voteInFrequency[\indexDigit][\indexActiveSymbol]}]=
	\sqrt{\symbolEnergy} \randomSymbolAtSubcarrier[\indexED,\indexGradient,\indexDigit]\times\indicatorFunction[{\digit[\indexED,\indexGradient][\indexDigit]=\symbol[\indexActiveSymbol]}]~,
	\label{eq:symbolPlus}
\end{align}
for all $\indexDigit\in\integers_\numberOfDigits$ and $\indexActiveSymbol\in\integers_{\base-1}$, where $\symbolEnergy$ is a factor to normalize the OFDM symbol energy and $\randomSymbolAtSubcarrier[\indexED,\indexGradient,\indexDigit]$ is a randomization symbol on the unit circle for \ac{PMEPR} reduction \cite{sahinCommnet_2021}. Note that we do not allocate a subcarrier for $\symbol[\base-1]=0$ as it does not contribute to the sum given in \eqref{eq:digitAveraging}. Since we
active only one of the $\base-1$ subcarriers in our scheme, we set $\symbolEnergy$ to $\base-1$.
 After the calculation of \eqref{eq:symbolPlus} for all gradients, the $\indexED$th \ac{ED} calculates the \ac{OFDM} symbols and all \acp{ED} transmit them simultaneously based on the discussions in Section~\ref{sec:system}. Since the proposed scheme uses $(\base-1)\numberOfDigits$ subcarriers for each gradient, the maximum number of gradients that can be transmitted on each \ac{OFDM} symbol can be calculated as $\numberOfParametersPerOFDM=\floor{\numberOfActiveSubcarriers/((\base-1)\numberOfDigits)}$ for all \acp{ED}. 

 It is worth emphasizing that the function $\mappingFunction$ can be designed based on a scrambler to randomize the synthesized \ac{OFDM} symbols or an encryption function to enhance the security of the \ac{OAC}. We leave these extensions for future work and assume that the function $\mappingFunction$ uses $(\base-1)\numberOfDigits$ adjacent subcarriers for each gradient, as illustrated in \figurename~\ref{fig:feelBlockDiagram}. In addition, we do not use \ac{TCI} to compensate for the multipath channel as this is beneficial to eliminate 1) the need for precise time synchronization, 2) the channel estimation overhead, 3) the information loss due to the truncation, and 4) the instantaneous power fluctuations in fading channel due to the channel inversion. Our scheme also relies on a non-coherent receiver as discussed in Section~\ref{subsec:edrec}.

 \begin{example}
 	\rm 
 	Consider the parameters given in Example~\ref{ex:sum}, i.e., $\numberOfEdgeDevices=2$, $\localGradientElement[0,\indexGradient][\indexCommunicationRound]=0.28$, and $\localGradientElement[1,\indexGradient][\indexCommunicationRound]=-0.86$, where the local gradients are represented as  $\encoder(0.28)=\representationInBase[\base][{\digit[0,\indexGradient][2],\digit[0,\indexGradient][1],\digit[0,\indexGradient][0]}]=\representationInBase[5][1,-2,2]$ for the $0$th \ac{ED}, and $\encoder(-0.86)=\representationInBase[\base][{\digit[1,\indexGradient][2],\digit[1,\indexGradient][1],\digit[1,\indexGradient][0]}]=\representationInBase[5][-2,-1,2]$  for the $1$st \ac{ED} for $\base=5$ and $\numberOfDigits=3$. Assume that the resource set for the $\indexGradient$th gradient, i.e., $\resourceSet[\indexGradient]$, is given by
 	\begin{align}
		\resourceSet[\indexGradient] = \{&
		(\voteInTime[0][0],\voteInFrequency[0][0]), 
		(\voteInTime[0][1],\voteInFrequency[0][1]), 
		(\voteInTime[0][2],\voteInFrequency[0][2]), 
		(\voteInTime[0][3],\voteInFrequency[0][3]), \nonumber\\	&
		(\voteInTime[1][0],\voteInFrequency[1][0]), 
		(\voteInTime[1][1],\voteInFrequency[1][1]),
		(\voteInTime[1][2],\voteInFrequency[1][2]), 
		(\voteInTime[1][3],\voteInFrequency[1][3]), \nonumber\\ &	
		(\voteInTime[2][0],\voteInFrequency[2][0]), 
		(\voteInTime[2][1],\voteInFrequency[2][1]), 
		(\voteInTime[2][2],\voteInFrequency[2][2]), 
		(\voteInTime[2][3],\voteInFrequency[2][3]), 			
		\} \nonumber\\
		=\{&
		(0,0),(0,1),\mydots,(0,11)	\}~, \nonumber
 	\end{align}
 	i.e., the first $12$ adjacent subcarriers of the $0$th \ac{OFDM} symbol. Based on \eqref{eq:setDef}, 
 	$\symbolSet[5]=\{\symbol[0]=-1,\symbol[1]=1,\symbol[2]=-2,\symbol[3]=2,\symbol[4]=0\}$.  Hence, based on \eqref{eq:symbolPlus},  the activated subcarriers for the $0$th \ac{ED} (with omitting the randomization symbols for readability) are given by 
 	\begin{align}
		(\transmittedSymbolAtSubcarrier[0,{0,0}],\mydots,\transmittedSymbolAtSubcarrier[0,{0,11}]) = (
\underbrace{0,0,0,\sqrt{\symbolEnergy}}_{\indexDigit=0},
\underbrace{0 ,0,\sqrt{\symbolEnergy},0}_{\indexDigit=1},
\underbrace{0,\sqrt{\symbolEnergy},0,0}_{\indexDigit=2}
)\nonumber~,
 	\end{align}
 as $\indicatorFunction[{\digit[0,\indexGradient][\indexDigit]=\symbol[\indexActiveSymbol]}]=1$ for $(\indexDigit=0,\indexActiveSymbol=3)$, $(\indexDigit=1,\indexActiveSymbol=2)$, and $(\indexDigit=2,\indexActiveSymbol=1)$. For the first \ac{ED}, the active subcarriers are given by
   	\begin{align}
  	(\transmittedSymbolAtSubcarrier[1,{0,0}],\mydots,\transmittedSymbolAtSubcarrier[1,{0,11}]) = (
  	\underbrace{0,0,0,\sqrt{\symbolEnergy}}_{\indexDigit=0},
  	\underbrace{\sqrt{\symbolEnergy} ,0,0,0}_{\indexDigit=1},
  	\underbrace{0,0,\sqrt{\symbolEnergy},0}_{\indexDigit=2}
  	)\nonumber~.
  \end{align}
as $\indicatorFunction[{\digit[0,\indexGradient][\indexDigit]=\symbol[\indexActiveSymbol]}]=1$ for $(\indexDigit=0,\indexActiveSymbol=0)$, $(\indexDigit=1,\indexActiveSymbol=2)$, and $(\indexDigit=2,\indexActiveSymbol=2)$.
 \end{example}
 
 \begin{remark}
 	\rm
For $\numberOfDigits=1$, the proposed scheme divides $[-\valueMaximum,\valueMaximum]$ into $\base-1$ equal intervals (or equivalently $[-\valueMaximumPrime,\valueMaximumPrime]$ into $\base$ intervals)  and the modulation is $(\base-1)$-ary \ac{FSK} over \ac{OFDM}, i.e., the scheme encodes the amplitude information into a subcarrier index via $\encoder$.
 \end{remark}

\subsection{Edge Server - Receiver}
\label{subsec:edrec}
\begin{figure*}[t]
	\centering
	{\includegraphics[width =\textwidth]{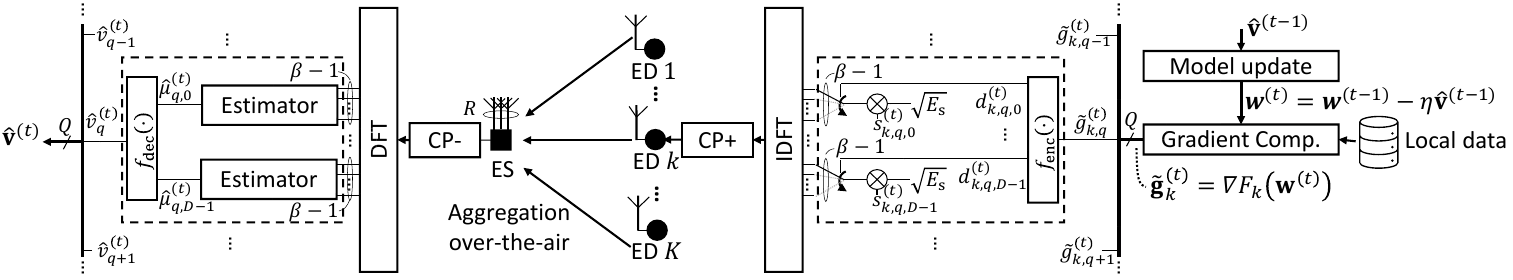}
	} 
	\caption{The transmitter and received diagrams with the proposed \ac{OAC} scheme for \ac{FEEL}.} 
	\label{fig:feelBlockDiagram}
\end{figure*}

 At the \ac{ES},  we  assume that the \ac{CSI}, i.e., $\{\channelAtSubcarrier[\indexED,\indexOFDMSymbol,\indexSubcarrier]|\indexED\in\integers_\numberOfEdgeDevices,\indexSubcarrier\in\integers_\numberOfActiveSubcarriers,\indexOFDMSymbol\in\integers_\numberOfOFDMSymbols\}$, is {\em not} available. Hence, the \ac{ES} exploits that
 $\receivedSymbolAtSubcarrier[{\voteInTime[\indexDigit][\indexActiveSymbol],\voteInFrequency[\indexDigit][\indexActiveSymbol]}]$ is a random vector for $\receivedSymbolAtSubcarrier[{\voteInTime[\indexDigit][\indexActiveSymbol],\voteInFrequency[\indexDigit][\indexActiveSymbol]}]\sim\complexGaussian[\zeroVector[\numberOfAntennasAtES]][{(\symbolEnergy\numberOFEDsForOptionGeneral[\indexActiveSymbol]+\noiseVariance)}{\identityMatrix[\numberOfAntennasAtES]}]$ and obtains an estimate of  $\{\numberOFEDsForOptionGeneral[\indexActiveSymbol]|\indexActiveSymbol\in\integers_{\base-1}\}$, non-coherently. For given $\indexDigit$ and $\indexGradient$, by using the corresponding log-likelihood function, the  \ac{ML} detector   can be expressed   as
\def\covarianceMatrix{{\bf \Sigma}_{\indexActiveSymbol}}
\def\aVector{{\rm \textbf{x}}_{\indexActiveSymbol}}
\begin{align}
	\{\numberOFEDsForOptionGeneralDetector[\indexActiveSymbol]|\indexActiveSymbol\in\integers_{\base-1}\}
	=&\arg\min_{\{\numberOFEDsForOptionGeneralVar[\indexActiveSymbol]\}}\left\{\sum_{\indexActiveSymbol=0}^{\base-2}\ln\det\covarianceMatrix+\aVector^{\rm H}\covarianceMatrix^{-1}\aVector\right\}	\label{eq:originalProblem}
	\\&
	\text{s.t.}~ 
	\sum_{\indexActiveSymbol=0}^{\base-2}\numberOFEDsForOptionGeneralVar[\indexActiveSymbol]\le\numberOfEdgeDevices, \numberOFEDsForOptionGeneralVar[\indexActiveSymbol]\in\{0,\mydots,\numberOfEdgeDevices\}, \forall\indexActiveSymbol\nonumber
\nonumber
\end{align}
where $\aVector = [\Re\{\receivedSymbolAtSubcarrier[{\voteInTime[\indexDigit][\indexActiveSymbol],\voteInFrequency[\indexDigit][\indexActiveSymbol]}]\}^{\rm T} ~\Im\{\receivedSymbolAtSubcarrier[{\voteInTime[\indexDigit][\indexActiveSymbol],\voteInFrequency[\indexDigit][\indexActiveSymbol]}]\}^{\rm T}]^{\rm T}$ and $\covarianceMatrix=\frac{\symbolEnergy\numberOFEDsForOptionGeneralVar[\indexActiveSymbol]+\noiseVariance}{2}\identityMatrix[2\numberOfAntennasAtES]$. However, due to the constraints, a solution to \eqref{eq:originalProblem} can increase the receiver complexity considerably. To address this issue, we relax the constraints and evaluate $\numberOFEDsForOptionGeneralDetector[\indexActiveSymbol]$ independently as given by
\begin{align}
	\numberOFEDsForOptionGeneralDetector[\indexActiveSymbol]&=\arg \min_{\numberOFEDsForOptionGeneralVar[\indexActiveSymbol]} \left\{2\numberOfAntennasAtES\ln\left(\frac{\symbolEnergy\numberOFEDsForOptionGeneralVar[\indexActiveSymbol]+\noiseVariance}{2}\right)+\frac{2\norm{\receivedSymbolAtSubcarrier[{\voteInTime[\indexDigit][\indexActiveSymbol],\voteInFrequency[\indexDigit][\indexActiveSymbol]}]}_2^2}{\symbolEnergy\numberOFEDsForOptionGeneralVar[\indexActiveSymbol]+\noiseVariance}\right\}\nonumber\\
	&=\frac{\norm{\receivedSymbolAtSubcarrier[{\voteInTime[\indexDigit][\indexActiveSymbol],\voteInFrequency[\indexDigit][\indexActiveSymbol]}]}_2^2}{\symbolEnergy\numberOfAntennasAtES}-\frac{\noiseVariance}{\symbolEnergy}~.
	\label{eq:edEst}
\end{align}
Therefore, a low-complexity estimator of $\digitAveraged[\indexGradient][\indexDigit]$   can be obtained as
\begin{align}
	\digitAveragedEst[\indexGradient][\indexDigit]=
	\frac{1}{\numberOfEdgeDevices}\sum_{\indexActiveSymbol=0}^{\base-2} \symbol[\indexActiveSymbol]\numberOFEDsForOptionGeneralDetector[\indexActiveSymbol]~.
	\label{eq:digitAveEst}
\end{align}
Finally, the estimator of $\meanGradientEle[\indexCommunicationRound][\indexGradient]$ can be expressed as
\begin{align}
	\meanGradientEleEstimate[\indexCommunicationRound][\indexGradient]=\decoder\representationInBase[\base][{\digitAveragedEst[\indexGradient][\numberOfDigits-1],\mydots,\digitAveragedEst[\indexGradient][1],\digitAveragedEst[\indexGradient][0]}]~.
	\label{eq:finalValEst}	
\end{align}
The \ac{ES} finally transmits $\meanGradientVectorEstimate[\indexCommunicationRound]$ to the \acp{ED} for the next communication round and the $\indexED$th \ac{ED} updates its parameters as $
	\modelParametersAtIteration[\indexCommunicationRound+1] = \modelParametersAtIteration[\indexCommunicationRound] - \learningRate  \meanGradientVectorEstimate[\indexCommunicationRound]
$, $\forall\indexED$. The corresponding transmitter and receiver diagrams  are provided in \figurename~\ref{fig:feelBlockDiagram}.


\subsection{MSE Analysis}
\label{subsec:mse}
For a given set of local stochastic gradients (i.e., given $\numberOFEDsForOptionGeneral[\indexActiveSymbol]$),
$\receivedSymbolAtSubcarrier[{\voteInTime[\indexDigit][\indexActiveSymbol],\voteInFrequency[\indexDigit][\indexActiveSymbol]}]\sim\complexGaussian[\zeroVector[\numberOfAntennasAtES]][{(\symbolEnergy\numberOFEDsForOptionGeneral[\indexActiveSymbol]+\noiseVariance)}{\identityMatrix[\numberOfAntennasAtES]}]$ holds for Rayleigh fading channel. Since the absolute square of an element of	$\receivedSymbolAtSubcarrier[{\voteInTime[\indexDigit][\indexActiveSymbol],\voteInFrequency[\indexDigit][\indexActiveSymbol]}]$ is an exponential distribution with the mean  $\symbolEnergy\numberOFEDsForOptionGeneral[\indexActiveSymbol]+\noiseVariance$,
the distribution of $\norm{\receivedSymbolAtSubcarrier[{\voteInTime[\indexDigit][\indexActiveSymbol],\voteInFrequency[\indexDigit][\indexActiveSymbol]}]}_2^2/\numberOfAntennasAtES$ can be obtained as  $\gammaDist[\numberOfAntennasAtES][{\numberOfAntennasAtES/(\symbolEnergy\numberOFEDsForOptionGeneral[\indexActiveSymbol]+\noiseVariance)}]$. 
As a result, the mean and the variance of the estimator $\numberOFEDsForOptionGeneralDetector[\indexActiveSymbol]$ can be calculated through the properties of a gamma distribution as
\begin{align}
\expectationOperator[{\numberOFEDsForOptionGeneralDetector[\indexActiveSymbol]}][{}]=\frac{\expectationOperator[\norm{\receivedSymbolAtSubcarrier[{\voteInTime[\indexDigit][\indexActiveSymbol],\voteInFrequency[\indexDigit][\indexActiveSymbol]}]}_2^2/\numberOfAntennasAtES][]}{\symbolEnergy}-\frac{\noiseVariance}{\symbolEnergy}=\numberOFEDsForOptionGeneral[\indexActiveSymbol]~,
\end{align}
and
\begin{align}
\varianceOperator[{\numberOFEDsForOptionGeneralDetector[\indexActiveSymbol]}]=\frac{\varianceOperator[\norm{\receivedSymbolAtSubcarrier[{\voteInTime[\indexDigit][\indexActiveSymbol],\voteInFrequency[\indexDigit][\indexActiveSymbol]}]}_2^2/\numberOfAntennasAtES]}{\symbolEnergy}=\frac{1}{\numberOfAntennasAtES}\left(\numberOFEDsForOptionGeneral[\indexActiveSymbol]+\frac{\noiseVariance}{\symbolEnergy}\right)^2~,
\label{eq:variance}
\end{align}
respectively, where the expectation is calculated over the randomness of the channel and noise. Hence, $\numberOFEDsForOptionGeneralDetector[\indexActiveSymbol]$ is an unbiased estimator. Also, based on \eqref{eq:digitAveEst} and \eqref{eq:finalValEst}, both $\digitAveragedEst[\indexGradient][\indexDigit]$ and $\meanGradientEleEstimate[\indexCommunicationRound][\indexGradient]$ are unbiased estimators of $\digitAveraged[\indexGradient][\indexDigit]$ and $\meanGradientEleOverQuantized[\indexCommunicationRound][\indexGradient]$, respectively. For a given  $\{\numberOFEDsForOptionGeneral[\indexActiveSymbol]| \indexActiveSymbol\in\integers_{\base-1}\}$, by using \eqref{eq:digitAveEst} and \eqref{eq:variance},  the variance of the estimator $\digitAveragedEst[\indexGradient][\indexDigit]$ is obtained as
\begin{align}
\varianceOperator[{{\digitAveragedEst[\indexGradient][\indexDigit]}}]=\frac{1}{\numberOfAntennasAtES\numberOfEdgeDevices^2}\sum_{\indexActiveSymbol=0}^{\base-2}\symbol[\indexActiveSymbol]^2\left(\numberOFEDsForOptionGeneral[\indexActiveSymbol]+\frac{\noiseVariance}{\symbolEnergy}\right)^2~.
\end{align} 
Therefore, we can calculate the variance of the estimator ${{\meanGradientEleEstimate[\indexCommunicationRound][\indexGradient]}}$ as
\begin{align}
	&\varianceOperator[{{\meanGradientEleEstimate[\indexCommunicationRound][\indexGradient]}}]=\frac{\valueMaximum^2}{\normalizationDigit^2\numberOfAntennasAtES\numberOfEdgeDevices^2}\sum_{\indexDigit=0}^{\numberOfDigits-1}\sum_{\indexActiveSymbol=0}^{\base-2}\symbol[\indexActiveSymbol]^2\left(\numberOFEDsForOptionGeneral[\indexActiveSymbol]+\frac{\noiseVariance}{\symbolEnergy}\right)^2\base^{2\indexDigit}~.
	\label{eq:varGivenK}
\end{align}
Hence, the (classical) \ac{MSE} of the estimator  $\meanGradientEleEstimate[\indexCommunicationRound][\indexGradient]$  can be obtained as
\begin{align}
	&\classicalMSEOperator[{\meanGradientEleEstimate[\indexCommunicationRound][\indexGradient]}]=\varianceOperator[{{\meanGradientEleEstimate[\indexCommunicationRound][\indexGradient]}}]+\frac{1}{\numberOfEdgeDevices^2}\left(\sum_{\indexED=0}^{\numberOfEdgeDevices-1} \localGradientElementQuantized[\indexED,\indexGradient][\indexCommunicationRound]-\localGradientElement[\indexED,\indexGradient][\indexCommunicationRound]\right)^2\nonumber~,
\end{align}	
where the last term  is the squared bias, i.e., $(\meanGradientEle[\indexCommunicationRound][\indexGradient]-\meanGradientEleOverQuantized[\indexCommunicationRound][\indexGradient])^2$, due to the quantization. 

The \ac{BMSE} of the estimator $\meanGradientEleEstimate[\indexCommunicationRound][\indexGradient]$ can be calculated as
\begin{align}
	\MSEOperator[{\meanGradientEleEstimate[\indexCommunicationRound][\indexGradient]}]&=\expectationOperator[{\classicalMSEOperator[{\meanGradientEleEstimate[\indexCommunicationRound][\indexGradient]}]}][]\nonumber\\&=\underbrace{\expectationOperator[{\left(\meanGradientEleEstimate[\indexCommunicationRound][\indexGradient]-\meanGradientEleOverQuantized[\indexCommunicationRound][\indexGradient]\right)^2}][{\meanGradientEleOverQuantized[\indexCommunicationRound][\indexGradient]}]}_{\coefficientOne}+\underbrace{\expectationOperator[{(\meanGradientEleOverQuantized[\indexCommunicationRound][\indexGradient]-\meanGradientEle[\indexCommunicationRound][\indexGradient])^2}][{\meanGradientEle[\indexCommunicationRound][\indexGradient]}]}_{\coefficientTwo}\nonumber
\end{align}
To derive the \ac{BMSE}, we assume that the distribution of $\localGradientElement[\indexED,\indexGradient][\indexCommunicationRound]$ is $\uniformDist[-\valueMaximumPrime][\valueMaximumPrime]$. 
Based on the derivation given in Appendix~\ref{app:derChannel}, $\coefficientOne$ can be calculated as
\begin{align}
	&\coefficientOne=\valueMaximum^2 		\underbrace{\frac{1}{3\numberOfAntennasAtES}\left(\frac{1}{\base} \left(1+\frac{\base\sigma_{\rm n}^2}{\numberOfEdgeDevices(\base-1)}\right)^2+\frac{\base}{\numberOfEdgeDevices(\base-1)}\right)\frac{\base^{\numberOfDigits}+1}{\base^\numberOfDigits-1}}_{\configurationFactorChannel}~.
	\label{eq:varChannel}
\end{align}
Since $\digit[\indexED,\indexGradient][\indexDigit]$ follows a uniform distribution for $\localGradientElement[\indexED,\indexGradient][\indexCommunicationRound]\sim\uniformDist[-\valueMaximumPrime][\valueMaximumPrime]$, we can obtain $\coefficientTwo$ as
\begin{align}
	\coefficientTwo=\valueMaximum^2\underbrace{\frac{1}{3\numberOfEdgeDevices(\base^\numberOfDigits-1)^2}}_{\configurationFactorQuan}
	~.\label{eq:quan}
\end{align} 
Therefore,  the \ac{BMSE}  can be calculated as
\begin{align}
	&\MSEOperator[{\meanGradientEleEstimate[\indexCommunicationRound][\indexGradient]}]
	=\coefficientOne+\coefficientTwo
=\valueMaximum^2\configurationFactorTotal~,
\label{eq:MSE}	
\end{align}	
where $\configurationFactorTotal$ is $\configurationFactorChannel+\configurationFactorQuan$.

In practice, the gradients often have an unknown probability distribution that changes over the communication rounds \cite{Zhang_2021}.  Hence, the expression in \eqref{eq:MSE} has limitations due to the underlying distribution assumption. On the other hand, the analysis with a general non-stationary distribution is much more complicated because the expected value in \eqref{eq:MSEpre} for different numerals may not be identical to each other. 
 Nevertheless, \eqref{eq:MSE} is a closed-form expression and predicts the performance of the scheme for a given configuration roughly without using sophisticated expressions, as exemplified in Section~\ref{sec:numerical}.
 
 Based on \eqref{eq:MSE}, we infer the followings: 1) The BMSE decreases with increasing $\base$ as both $\configurationFactorChannel$ and $\configurationFactorQuan$ tend to be smaller with a larger  $\base$. While increasing the number of numerals $\numberOfDigits$ decreases the factor $\configurationFactorQuan$, its impact on the factor $\configurationFactorChannel$ is limited as the limit of ${\base^{\numberOfDigits}+1}/({\base^\numberOfDigits-1})$ is $1$ as $\numberOfDigits$ approaches infinity. 2) BMSE decreases with the number of antennas in cases where the impact of the quantization error on the error is small for a larger $\base$ or a larger $\numberOfDigits$. 3) The impact of the quantization error on the BMSE rapidly diminishes  either by increasing $\base$ or $\numberOfDigits$.	4) The impact of $\noiseVariance$ on the \ac{BMSE}  decreases with the increasing number of \acp{ED}. 5) With increasing $\numberOfEdgeDevices$ or   $\base^\numberOfDigits$, the \ac{BMSE} asymptotically decreases   to $\valueMaximum^2/(3\numberOfAntennasAtES\base)$.

As we show in Section~\ref{sec:conv} and demonstrate in Section~\ref{sec:numerical},  quantization error plays a major role in the convergence rate of the \ac{FEEL}. To reduce quantization error over the communication rounds of \ac{FEEL}, we introduce a simple method in the following subsection.

\subsection{Adaptive Absolute Maximum (AAM)}
\def\metric[#1][#2]{m_{#1}^{(#2)}}
\def\metricVector[#1]{\textbf{m}^{(#1)}}
\def\metricFactor{\alpha}
Without any adaptation, the BMSE in \eqref{eq:MSE} is a constant, and the error due to the proposed scheme can  dominate the estimate of $\meanGradientEle[\indexCommunicationRound][\indexGradient]$ when its value is closer to $0$. This can be a non-negligible issue in practice because the gradients tend to become smaller over time. To address this issue, we exploit the fact that the gradients between adjacent communication rounds may have a high correlation \cite{liang2021wynerziv} and  propose to improve the proposed scheme with a feedback loop where all the \acp{ED} transmit only a {\em single} parameter related to their local gradients to the \ac{ES} through a control channel (e.g., \ac{PUCCH} in 3GPP \ac{5G} \ac{NR}) and the \ac{ES} sets up a new  absolute maximum $\valueMaximum$ for the next communication round based on the received feedback from the \acp{ED}. The information that is transmitted from \ac{ED} can be a function of the maximum absolute value of the gradients, the empirical variance, the standard deviation, or the mean of the gradients. In this study, we assume that the feedback loop realizes the \ac{AAM} as
\begin{align}
	\valueMaximum^{(\indexCommunicationRound)}=\metricFactor\times\norm{\metricVector[\indexCommunicationRound-1]}_\infty~,
	\label{eq:aam}
\end{align}
where $\metricVector[\indexCommunicationRound]=[\metric[0][\indexCommunicationRound],\mydots,\metric[\numberOfEdgeDevices-1][\indexCommunicationRound]]$ is the metric vector, $\metric[\indexED][\indexCommunicationRound]$ is the metric for the $\indexED$th \ac{ED}, $\forall\indexED$,  $\metricFactor$ is a positive value, and $\valueMaximum^{(0)}$ is the initial value for the \ac{AAM}. 
The \ac{AAM} based on \eqref{eq:aam} can be implemented in a practical network as follows: 1) The $\indexED$th \ac{ED} transmits $\metric[\indexED][\indexCommunicationRound]$, $\forall\indexED$, at the $\indexCommunicationRound$th communication round through an orthogonal channel. 2) The \ac{ES} calculates \eqref{eq:aam}. 3) The \ac{ES} transmits $\valueMaximum^{(\indexCommunicationRound+1)}$ to the \acp{ED}. 4) The \acp{ED} update $\encoder$ with the new absolute maximum $\valueMaximum^{(\indexCommunicationRound+1)}$.

In this study, we choose $\metric[\indexED][\indexCommunicationRound]=\norm{\localGradient[\indexED][\indexCommunicationRound]}_2$ and $\metricFactor=5/\sqrt{\numberOfModelParameters}$, heuristically, based on five-sigma deviation rule. The convergence rate of \ac{FEEL} with and without \ac{AAM} is analyzed in Section~\ref{sec:conv}.

\section{Convergence Analysis}
\label{sec:conv}

For the convergence rate analysis, we consider well-known Lipschitz continuity \cite{nesterov2004introductory} and make several assumptions on the loss function and gradient estimates, given as follows:
\begin{definition}  \rm
	A function $\aFunction$ is  $\LipschitzConstant$-Lipschitz over a set $S$ with respect to a norm $\norm{\cdot}$ if there exists a real constant $\LipschitzConstant>0$ such that $\norm{\aFunction(\vectorY)-\aFunction(\vectorX)}\le\LipschitzConstant\norm{\vectorY-\vectorX}$, $\forall\vectorX,\vectorY\in S$.
\end{definition}
\begin{lemma}[{\cite[Lemma 1.2.3]{nesterov2004introductory}}] \rm
	For a differentiable function $\aFunction:\realNumbers^{\numberOfModelParameters}\rightarrow\realNumbers$, let $\nabla\aFunction$  be  $\LipschitzConstant$-Lipschitz on $\realNumbers^{\numberOfModelParameters}$ with respect to norm $\norm{\cdot}_2$. Then, for any $\vectorY,\vectorX$ from $\realNumbers^\numberOfModelParameters$,
	\begin{align}
		\left| \aFunction(\vectorY) - \aFunction(\vectorX)-\nabla\aFunction(\vectorX)^{\rm T}(\vectorY-\vectorX) \right| \le \frac{\LipschitzConstant}{2}\norm{\vectorY-\vectorX}_2^2~.
	\end{align}	
\end{lemma}
\begin{assumption}[Bounded loss function]
	\rm The loss function is bounded, i.e.,
	$\lossFunctionGlobal[\modelParameters]\ge \lossFunctionGlobalMinimum$, $\forall\modelParameters$. 
\label{assump:boundedLoss}	
\end{assumption}
\begin{assumption}[Smooth gradients] 
	\rm The gradient of the loss function, i.e.,
	$\gradientlossFunctionGlobal$,  is  $\LipschitzConstant$-Lipschitz on $\realNumbers^{\numberOfModelParameters}$ with respect to the norm $\norm{\cdot}_2$, i.e., $
\norm{\nabla\lossFunctionGlobal[\modelParameters']-\nabla\lossFunctionGlobal[\modelParameters]}_2\le\LipschitzConstant\norm{\modelParameters'-\modelParameters}_2, \forall\modelParameters,\modelParameters'\in \realNumbers^{\numberOfModelParameters}$.
\label{assump:smooth}
\end{assumption}

\begin{figure}[t]
	\centering
	{\includegraphics[width =3.0in]{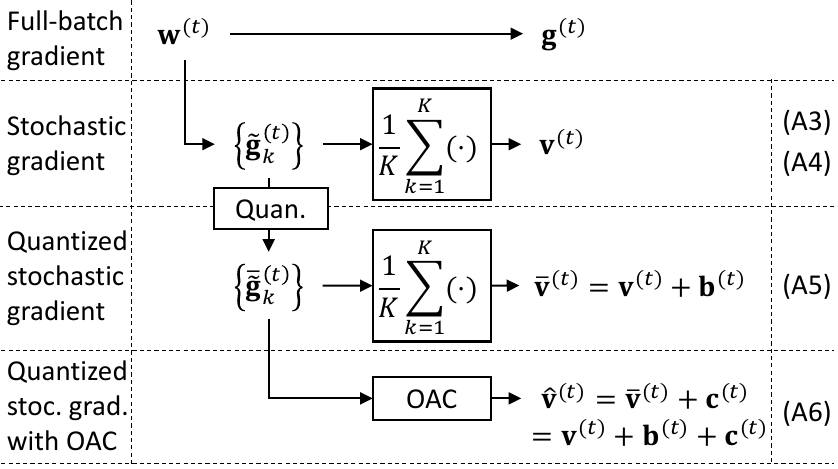}
	} 
	\caption{Relationships between the variables and the assumptions.  Assumptions \ref{assump:unbiasAverage}-\ref{assump:mse} are denoted as A3-A6, respectively.}
	\label{fig:relationships}
\end{figure}

Assumption 1 and Assumption 2 are the standard assumptions that are often made in the literature for convergence analysis.

\begin{assumption}[Unbiased average local stochastic gradients]
	\rm The average stochastic  gradient vector is an unbiased estimate of the global gradient vector, i.e., $		\expectationOperator[{\meanGradientVector[\indexCommunicationRound]}][] = \globalGradient[\indexCommunicationRound]$.
\label{assump:unbiasAverage}
\end{assumption}
\begin{assumption}[Gradient divergence]
	\rm For all ${\modelParametersAtIteration[\indexCommunicationRound]}$, the second-order moments of the local stochastic  gradients of the $\indexED$th \ac{ED}  with respected to the global gradients is bounded as $
	\expectationOperator[{\norm{\localGradient[\indexED][\indexCommunicationRound]-\globalGradient[\indexCommunicationRound]}_2^2}][]\le\secondMoment[\indexED],~\forall\indexED$. {This assumption implies that $\expectationOperator[\norm{\meanGradientVector[\indexCommunicationRound]-\globalGradient[\indexCommunicationRound]}_2^2][{}]\le\frac{1}{\numberOfEdgeDevices}\sum_{\indexED=0}^{\numberOfEdgeDevices-1}\secondMoment[\indexDigit]$.}
\label{assump:graDiv}
\end{assumption}

Assumption 3 and Assumption 4 do not require the local gradients to be unbiased estimates of the global gradients. Hence, they are compatible with a heterogeneous data distribution scenario where the {\em sum} of local gradients is unbiased.

\begin{assumption} [Average quantization bias]
	\rm Let $\biasQuan[\indexCommunicationRound]\triangleq{\meanGradientOverQuantized[\indexCommunicationRound]-\meanGradientVector[\indexCommunicationRound]}=\frac{1}{\numberOfEdgeDevices}\sum_{\indexED=1}^{\numberOfEdgeDevices}\localGradientVectorQuantized[\indexED][\indexCommunicationRound]-\localGradient[\indexED][\indexCommunicationRound]$ be the quantization bias averaged across the EDs. The expected value of the average quantization bias is zero, i.e., $\expectationOperator[{\biasQuan[\indexCommunicationRound]}][]=\zeroVector[\numberOfModelParameters]$.
\label{assump:aveQaun}
\end{assumption}
Assumption 5 gets weaker with increasing $\numberOfEdgeDevices$ due to the averaging across the EDs or  reducing the quantization step by increasing $\base$ or $\numberOfDigits$. 

\begin{assumption}[MSE bound]
	\rm 
	Let us express the aggregated gradient with OAC as $\meanGradientVectorEstimate[\indexCommunicationRound]=\meanGradientVector[\indexCommunicationRound]+\biasQuan[\indexCommunicationRound]+\channelNoise[\indexCommunicationRound]$, where $\channelNoise[\indexCommunicationRound]$ is the noise due to the OAC. The average \ac{MSE} due to the communication channel and the quantization is bounded by $
\expectationOperator[\norm{{\meanGradientVectorEstimate[\indexCommunicationRound]}-\meanGradientVector[\indexCommunicationRound]}^2_2][]\le\numberOfModelParameters(\coefficientOne+\coefficientTwo)$.
\label{assump:mse}
\end{assumption}
It is worth noting that the expected value of the channel noise is zero, i.e., $\expectationOperator[{\channelNoise[\indexCommunicationRound]}][]=\zeroVector[\numberOfModelParameters]$ since $\meanGradientEleEstimate[\indexCommunicationRound][\indexGradient]$ is an unbiased estimator  $\meanGradientEleOverQuantized[\indexCommunicationRound][\indexGradient]$. 
The relationship between the variables and the assumptions are given in \figurename~\ref{fig:relationships} for clarity.

%

\begin{theorem}
	\rm For a fixed learning rate $\learningRate$, the convergence rate of the distributed training based on the proposed scheme in the Rayleigh  channel is
	\begin{align}
	\expectationOperator[\frac{1}{\communicationRounds}\sum_{\indexCommunicationRound=0}^{\communicationRounds-1} \norm{\globalGradient[\indexCommunicationRound]}_2^2][]\nonumber\le&\frac{1}{{\communicationRounds\learningRate}\left(1-\frac{\learningRate\LipschitzConstant}{2}\right)} \left(	\lossFunctionGlobal[{\modelParametersAtIteration[0]}]- \lossFunctionGlobalMinimum\right)\nonumber\\&\hspace{-16mm}+\frac{\frac{\learningRate\LipschitzConstant}{2}}{1-\frac{\learningRate\LipschitzConstant}{2}}\left((\coefficientOne+\coefficientTwo)\numberOfModelParameters+\frac{1}{\numberOfEdgeDevices}\sum_{\indexED=0}^{\numberOfEdgeDevices-1}\secondMoment[\indexED]\right)~,
	\label{eq:convergence}
\end{align}
	where $\coefficientOne$ and $\coefficientTwo$ are given in \eqref{eq:varChannel} and \eqref{eq:quan}, respectively.
	\label{th:convergence}
\end{theorem}
The proof is given in Appendix~\ref{app:proofTh1}.

Theorem~\ref{th:convergence} is an extension of the convergence analysis of \ac{SGD} under the consideration of  the proposed scheme.
While the first term of the bound given in \eqref{eq:convergence} becomes smaller for a larger total number of communication rounds $\communicationRounds$, the noise ball is determined with the values of the learning rate $\learningRate$, the noise variance due to the local stochastic gradient estimates, and the noise due to the proposed scheme. The noise ball decreases when a smaller learning rate $\learningRate$ is used at the expense of a larger $\communicationRounds$ due to the first term in \eqref{eq:convergence}.
The proposed scheme contributes to the noise variance due to stochastic gradient calculation in \eqref{eq:sgd}. Hence, the standard tuning methods for \ac{SGD} such as momentum can also be utilized with the proposed scheme to improve the convergence rate.

The convergence rate of the \ac{FEEL} under the presence of the proposed scheme with \ac{AAM} based on \eqref{eq:aam} can be expressed as follows:
\def\newLipschitz{L'}
\begin{theorem}
	\rm For a fixed learning rate $\learningRate$, the convergence rate of the distributed training based on the proposed scheme with AAM in the Rayleigh channel is
\begin{align}
	\expectationOperator[\frac{1}{\communicationRounds}\sum_{\indexCommunicationRound=1}^{\communicationRounds} \norm{\globalGradient[\indexCommunicationRound]}_2^2][]\nonumber&\le\frac{1}{{\communicationRounds\learningRate}\left(1-\frac{\learningRate\newLipschitz}{2}\right)} \left(	\lossFunctionGlobal[{\modelParametersAtIteration[1]}]- \lossFunctionGlobalMinimum\right.
	\\&\left.~+\frac{\learningRate\LipschitzConstant}{2}\metricFactor^2\configurationFactorTotal\numberOfEdgeDevices\expectationOperator[{\norm{\globalGradient[0]}_2^2-\norm{\globalGradient[T]}_2^2}][]\right)\nonumber
	\\&~~+\frac{\frac{\learningRate\newLipschitz}{2}}{1-\frac{\learningRate\newLipschitz}{2}}\frac{1}{\numberOfEdgeDevices}\sum_{\indexED=0}^{\numberOfEdgeDevices-1}\secondMoment[\indexED]~,
	\label{eq:convergenceAAM}
\end{align}
	where $\newLipschitz=\LipschitzConstant(1+\metricFactor^2\configurationFactorTotal\numberOfEdgeDevices)$  for  $\numberOFEDsForOptionGeneral[\indexActiveSymbol]\sim\binomDist[\numberOfEdgeDevices][1/\base]$ for all $\indexActiveSymbol,\indexDigit,\indexGradient$.
	\label{th:convergenceAAM}
\end{theorem}
The proof is given in Appendix~\ref{app:proofTh2}.

Theorem~\eqref{th:convergenceAAM} shows that the \ac{AAM} eliminates the additive impact of the proposed scheme on the noise on the gradients (as in Theorem~\ref{th:convergence}) at the expense of scaling up the constant $\LipschitzConstant$. As compared to the case without \ac{AAM}, the noisy ball is smaller with \ac{AAM}. Hence, the convergence rate improves considerably, as demonstrated in Section~\ref{sec:numerical}.

\section{Numerical Results}
\label{sec:numerical}
In this section, we assess the proposed scheme numerically for  $\numberOfDigits\in\{1,2\}$  and $\base\in\{3,5,7\}$. We demonstrate its \ac{BMSE} performance and the test accuracy results based on \ac{FEEL} under homogeneous and heterogeneous data distributions. For the comparisons, we consider  Goldenbaum's scheme (without channel inversion) \cite{Goldenbaum_2013tcom,Goldenbaum_wcl2014} and \ac{FSK-MV} \cite{sahinCommnet_2021,Sahin_2022MVjournal} since they rely on non-coherent techniques, similar to the proposed scheme. We also provide the results without OAC based on \ac{SGD}. We do not consider methods based on channel inversion techniques as their performance can deteriorate quickly in the presence of synchronization errors \cite{sahinWCNC_2022,sahinCommnet_2021}. 

\def\coefA{a}
\def\coefB{b}
\def\varArb{x}
\def\sequenceLength{L}
\def\affineEnc{\epsilon}
\def\affineDec{\delta}
Goldenbaum's scheme aims to compute continuous-valued functions based on analog modulation. After the symbol at the $\indexED$th ED is processed with a function $\affineEnc(\varArb)=\coefA\varArb+\coefB$ that results in a non-negative value for $\coefA=1/\valueMaximum$ and $\coefB=\valueMaximum$, the square root of the resulting value is multiplied with a unimodular sequence of length $\sequenceLength$ as $\sqrt{\affineEnc(\localGradientElement[\indexED,\indexGradient][\indexCommunicationRound])}\times[
{\constante^{\constantj\theta_{\indexED,1}}} ,\mydots, {\constante^{\constantj\theta_{\indexED,\sequenceLength}}}]$. At the receiver, an estimate of the aggregated symbol is obtained after processing  the average energy of the received sequence across $\numberOfAntennasAtES$ antennas with another affine function $\affineDec(\varArb)=(\varArb-\numberOfEdgeDevices\coefB)/\coefA$ to reverse the impact of $\affineEnc(\varArb)$ on the superposed symbols. The main shortcoming of this scheme is that it causes additional interference terms for $\sequenceLength<\numberOfEdgeDevices$. For the numerical analysis, we consider $\sequenceLength\in\{4,12\}$ and choose the sequence elements from $\{1,-1,\constantj,-\constantj\}$ randomly. The EDs transmit the sequences by mapping them to the OFDM subcarriers.

\ac{FSK-MV} relies on digital modulation to represent two discrete states, i.e., $\{-1,1\}$, and targets to compute a specific function, i.e., \ac{MV}, for distributed training by \ac{MV}. With this method, even if the value of the symbol is very close to $0$, the transmitted values are $1$ and $-1$. Hence,  without any precaution, it can bias the training for the scenarios with heterogeneous data distribution.

For all simulations, we consider a single cell with $\numberOfEdgeDevices=25$ \acp{ED}. We set the \ac{SNR}, i.e., $1/\noiseVariance$, to be $20$~dB, and choose the number of antennas at the \ac{ES} as $\numberOfAntennasAtES\in\{1,25\}$.

\def\figuresize{\textwidth/2}
\begin{figure*}
	\centering
	\subfloat[Uniform distribution. The curves with the marker '$+$' and the line '-' are for the simulation and the theoretical  results for the proposed method, respectively.]{\includegraphics[width =\figuresize]{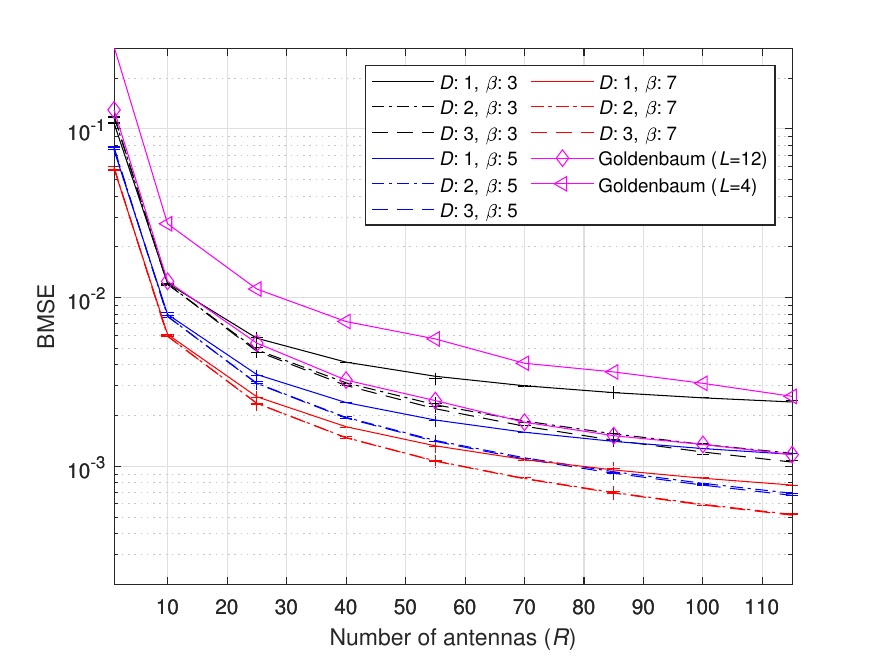}
		\label{subfig:mse_uni}}	
	\subfloat[Gaussian distribution (Simulation).]{\includegraphics[width =\figuresize]{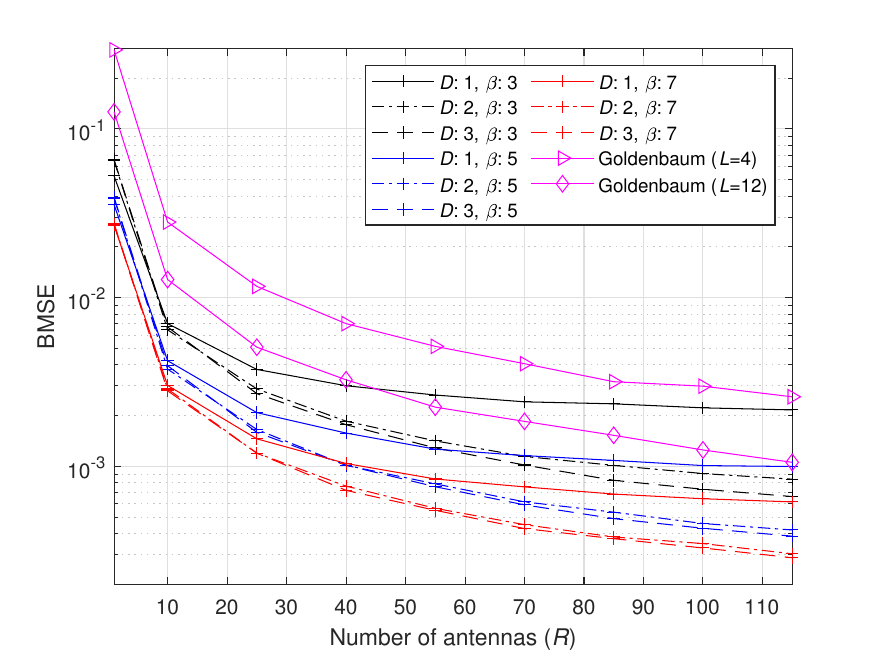}
		\label{subfig:mse_gau}}
	\caption{BMSE  versus the number of antennas for  $\localGradientElement[\indexED,\indexGradient][\indexCommunicationRound]\sim\uniformDist[-1][1]$ and $\localGradientElement[\indexED,\indexGradient][\indexCommunicationRound]\sim\gaussianDist[0][0.2]$, $\forall\indexED$ ($\noiseVariance=0.01$, $\numberOfEdgeDevices=25$ EDs). The proposed scheme provides less error for increasing $\base$ and $\numberOfDigits$.}
	\label{fig:MSE}	
\end{figure*}

\subsection{BMSE and error distribution}
In this subsection, we analyze the BMSE of the estimator of \eqref{eq:finalValEst}.  We calculate the BMSE through simulations for $\localGradientElement[\indexED,\indexGradient][\indexCommunicationRound]\sim\uniformDist[-1][1]$ and $\localGradientElement[\indexED,\indexGradient][\indexCommunicationRound]\sim\gaussianDist[0][0.2]$, $\forall\indexED$. For the proposed scheme, we set $\valueMaximum$ to $(\base^\numberOfDigits-1)/\base^\numberOfDigits$ by taking the quantization step into account. For Goldenbaum's scheme, we clamp the outcome if it is not within the range $[1,1]$ and set $\valueMaximum$ to $1$.

In  \figurename~\ref{fig:MSE}\subref{subfig:mse_uni} and \figurename~\ref{fig:MSE}\subref{subfig:mse_gau}, we plot the \ac{BMSE} versus the number of antennas for the uniform and Gaussian distributions, respectively. As can be seen from \figurename~\ref{fig:MSE}\subref{subfig:mse_uni}, the simulation results exactly match the theoretical results based on \eqref{eq:MSE}. The results are also aligned with the discussions provided in Section~\ref{subsec:mse}. Increasing $\base$ reduces the \ac{BMSE}. While a larger $\numberOfDigits$ decreases the \ac{BMSE} (by reducing the quantization error), its impact on the \ac{BMSE} quickly saturates. Similarly, Goldenbaum's scheme performance improves by increasing the number of antennas. However, its performance is slightly worse than the proposed scheme for the same amount of resource consumption. Similar observations can also be made from \figurename~\ref{fig:MSE}\subref{subfig:mse_gau} although the distribution is different from the uniform distribution. We also observe that the theoretical BMSE results are more pessimistic than the ones in this scenario. For example, the BMSE results for the uniform and Gaussian distribution for a single antenna are around $0.1$ and $0.07$, respectively. 

\begin{figure}[t]
	\centering
	{\includegraphics[width =3.5in]{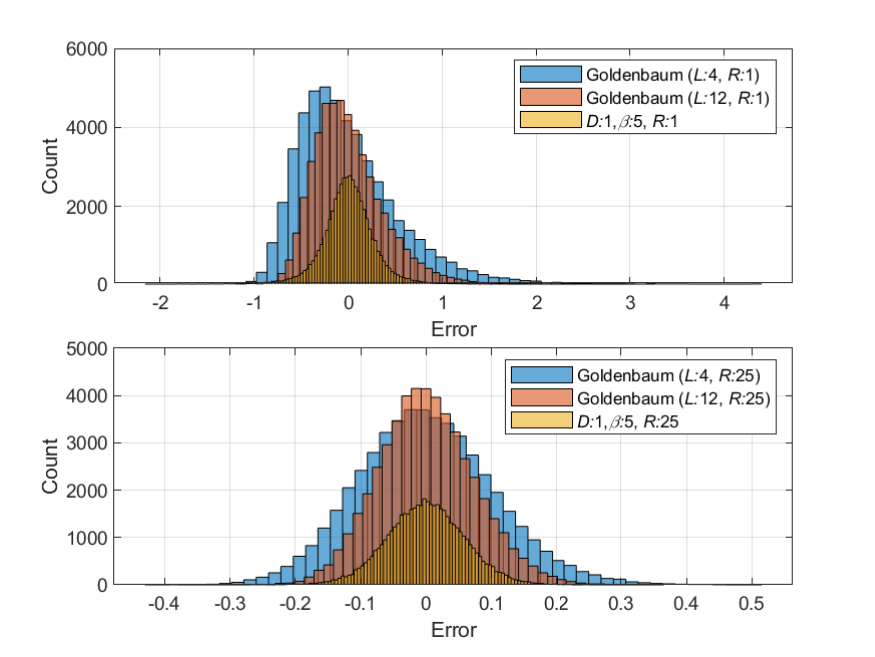}
	} 
	\caption{The error histogram for $\localGradientElement[\indexED,\indexGradient][\indexCommunicationRound]\sim\uniformDist[-1][1]$ ($\noiseVariance=0.01$, $\numberOfEdgeDevices=25$, 50000 realizations). While the error distribution for Goldenbaum's scheme is skewed for $\numberOfAntennasAtES=1$ antenna, it is symmetric around $0$ for the proposed scheme.}
	\label{fig:histogram}
\end{figure}
In \figurename~\ref{fig:histogram}, we plot the histogram of the error for the uniform distribution. The main observation is that the error distribution for Goldenbaum's scheme is  skewed for $\numberOfAntennasAtES=1$ antenna, while it is symmetric for the proposed scheme. For $\numberOfAntennasAtES=25$ antennas, Goldenbaum's scheme  becomes less skewed due to the channel hardening.

\subsection{FEEL}
\def\figuresize{3.0in}
\begin{figure*}
	\centering
	\subfloat[Without AAM (Momentum: 0,  $\valueMaximum=1$).]{\includegraphics[width =\figuresize]{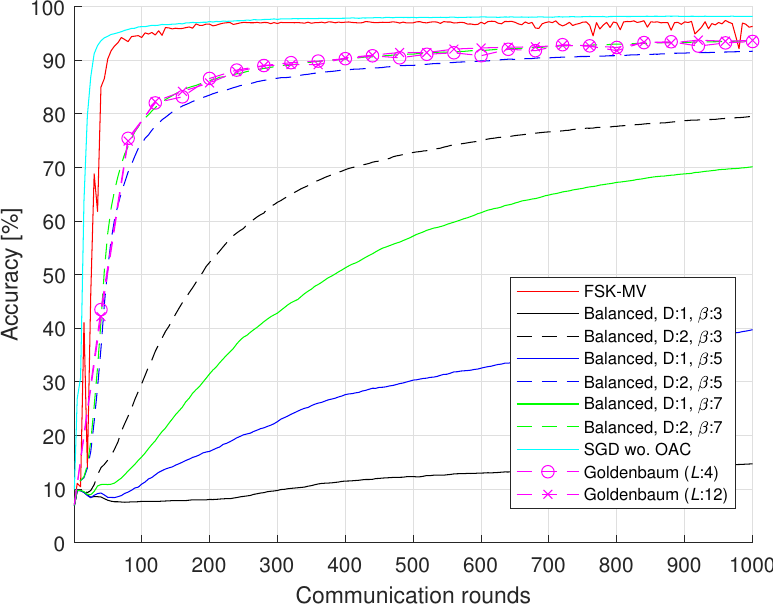}
		\label{subfig:acc_homo_R1_woM_woAAM}}~~~
	\subfloat[Without AAM (Momentum: 0.9,  $\valueMaximum=1$).]{\includegraphics[width =\figuresize]{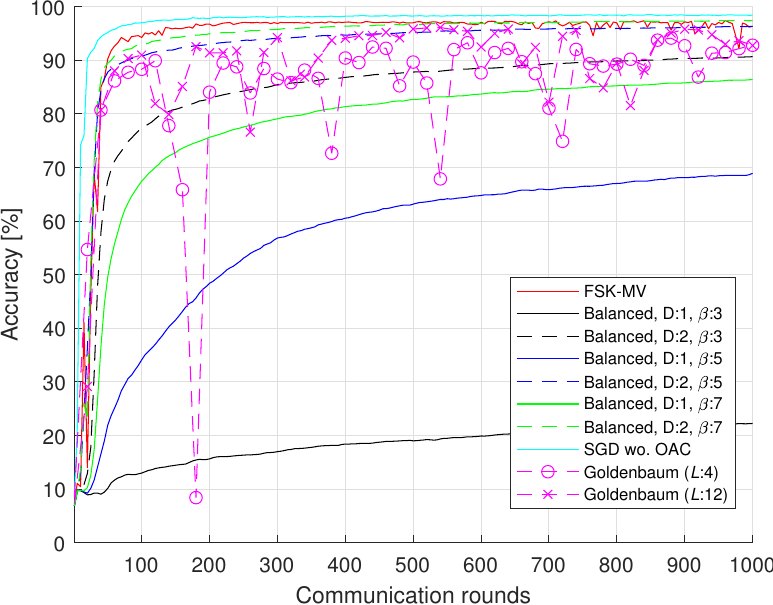}
		\label{subfig:acc_homo_R1_wM_w0AAM}}\\		
	\subfloat[With AAM (Momentum: 0).]{\includegraphics[width =\figuresize]{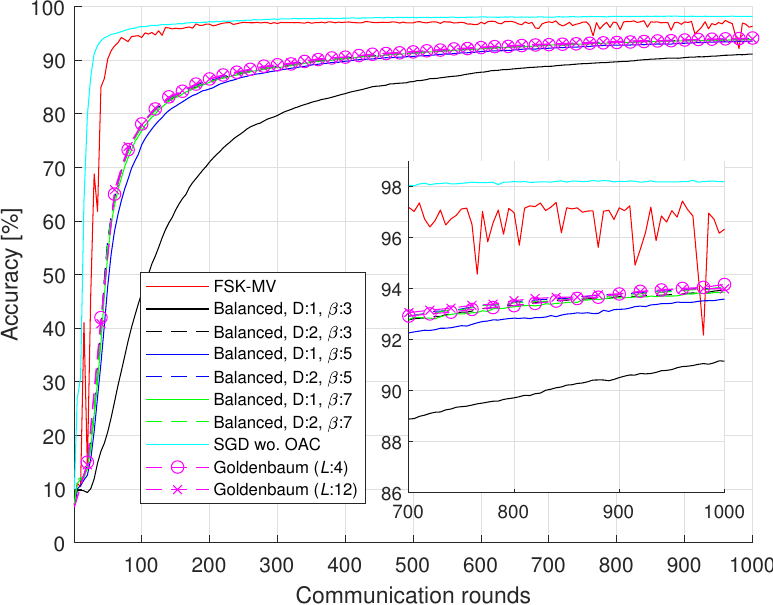}
		\label{subfig:acc_homo_R1_woM_wAAM}}~~~
	\subfloat[ With AAM (Momentum: 0.9).]{\includegraphics[width =\figuresize]{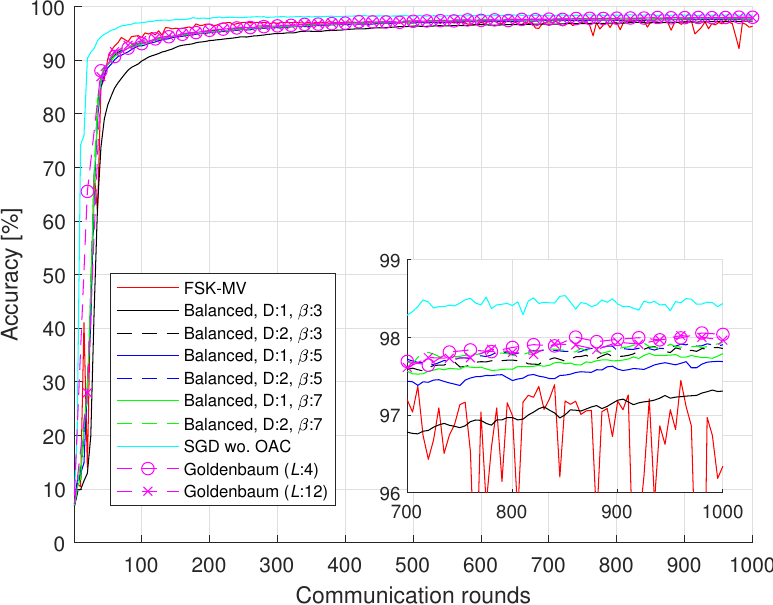}
		\label{subfig:acc_homo_R1_wM_wAAM}}
	\caption{Test accuracy versus communication rounds (Homogeneous data distribution, $\numberOfAntennasAtES=1$, $\numberOfEdgeDevices=25$). The proposed scheme with AAM addresses the quantization errors.}
	\label{fig:testAccHomo}
\end{figure*}
\begin{figure*}
	\centering
	\subfloat[Without AAM (Momentum: 0,  $\valueMaximum=1$).]{\includegraphics[width =\figuresize]{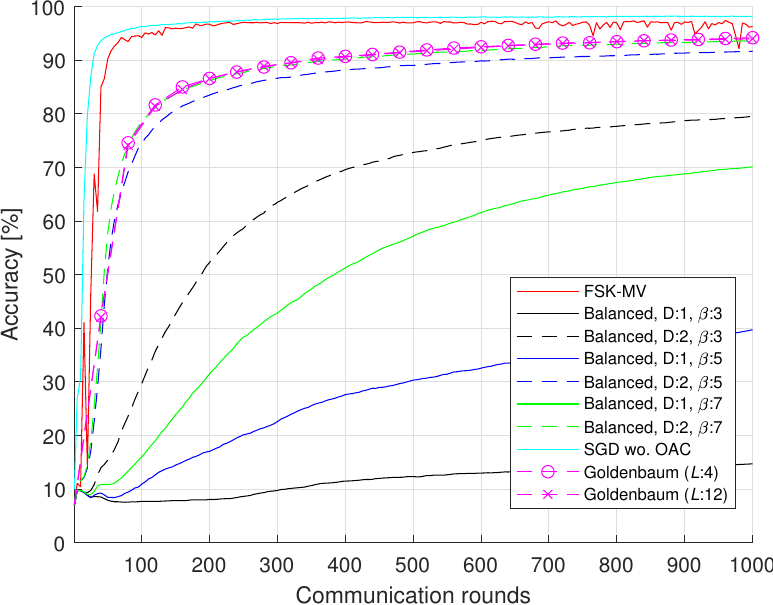}
		\label{subfig:acc_homo_R25_woM_woAAM}}~~~		
	\subfloat[Without AAM (Momentum: 0.9,  $\valueMaximum=1$).]{\includegraphics[width =\figuresize]{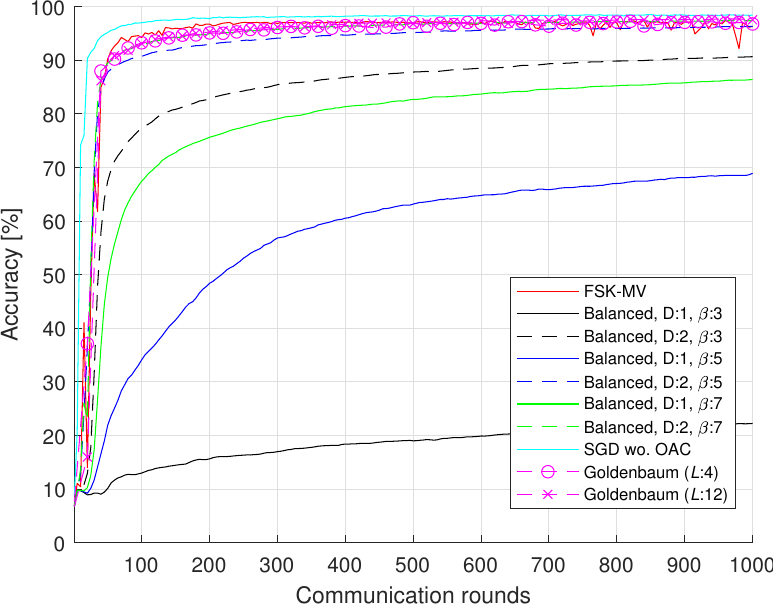}
		\label{subfig:acc_homo_R25_wM_woAAM}}\\			
	\subfloat[ With AAM (Momentum: 0).]{\includegraphics[width =\figuresize]{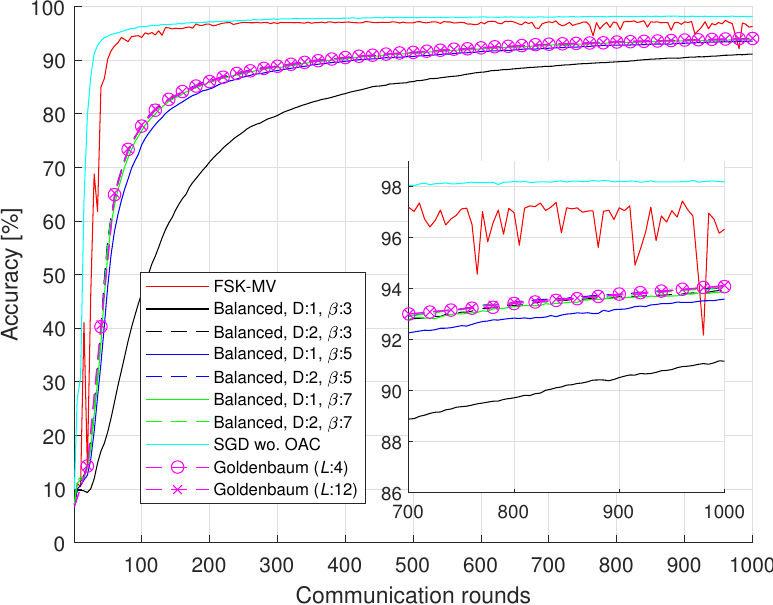}
		\label{subfig:acc_homo_R25_woM_wAAM}}~~~
	\subfloat[ With AAM (Momentum: 0.9).]{\includegraphics[width =\figuresize]{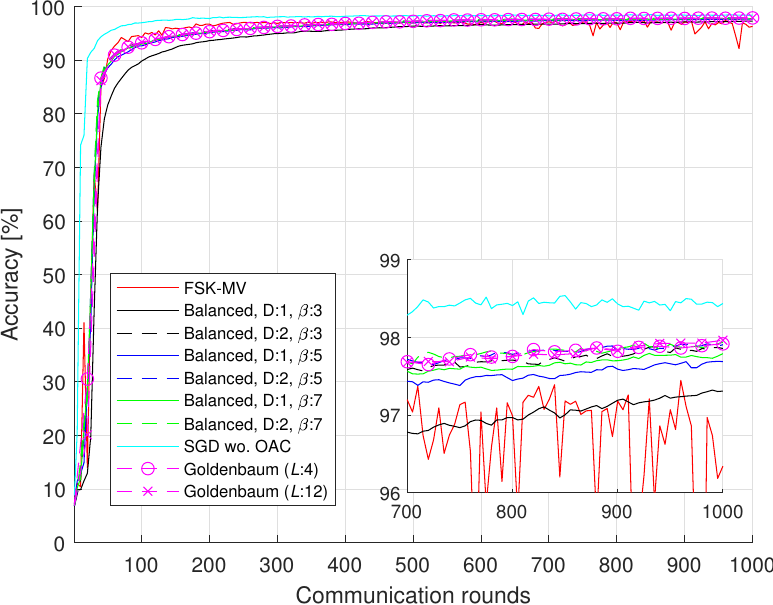}
		\label{subfig:acc_homo_R25_wM_wAAM}}		
	\caption{ Test accuracy versus communication rounds (Homogeneous data distribution, $\numberOfAntennasAtES=25$, $\numberOfEdgeDevices=25$). Increasing the number of antennas has a negligible effect on the test accuracy for the proposed scheme.}
	\label{fig:testAccHomoMoreAntennas}
\end{figure*}
\begin{figure*}
	\centering
	\subfloat[ Without AAM (Momentum: 0,  $\valueMaximum=1$).]{\includegraphics[width =\figuresize]{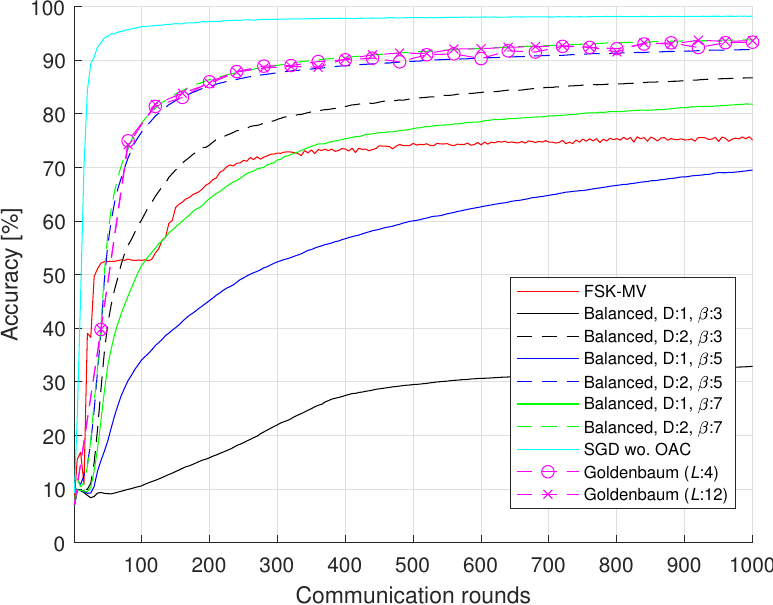}
		\label{subfig:acc_heto_R1_woM_woAAM}}~~~	
	\subfloat[ Without AAM (Momentum: 0.9,  $\valueMaximum=1$).]{\includegraphics[width =\figuresize]{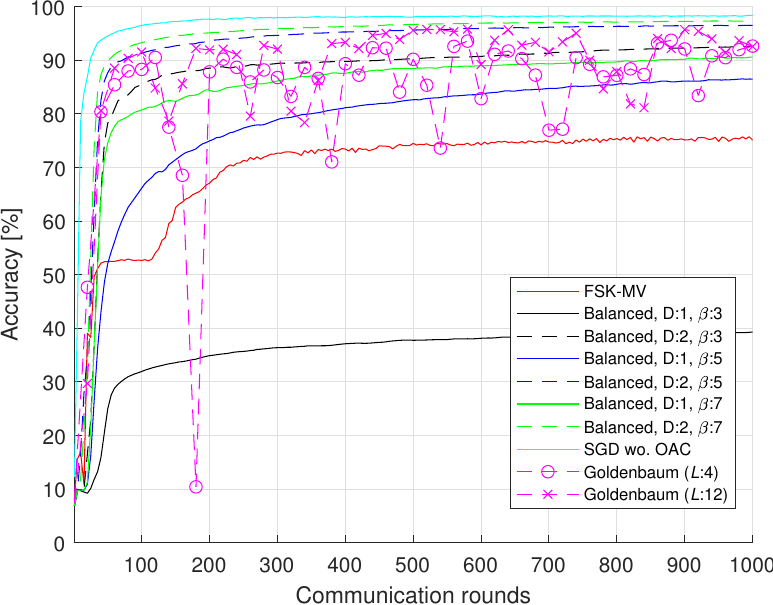}
		\label{subfig:acc_heto_R1_wM_woAAM}}\\		
	\subfloat[ With AAM (Momentum: 0).]{\includegraphics[width =\figuresize]{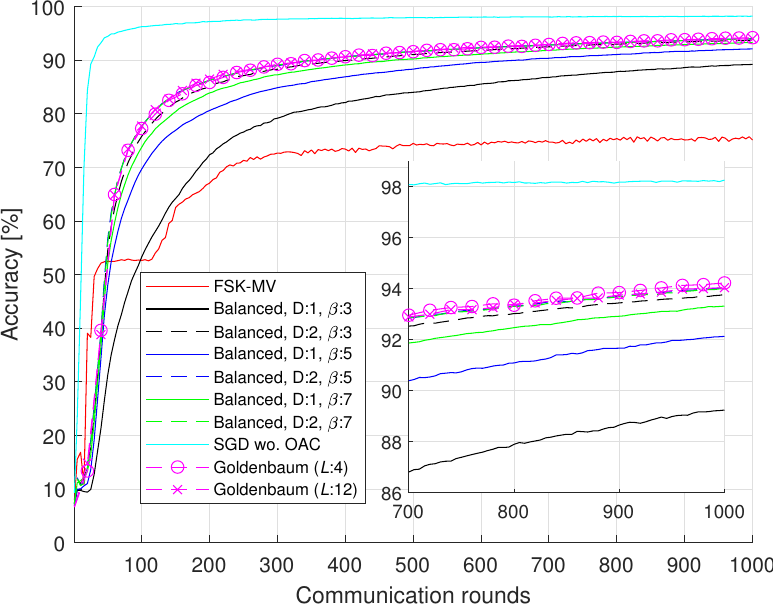}
		\label{subfig:acc_heto_R1_woM_wAAM}}~~~
	\subfloat[ With AAM (Momentum: 0.9).]{\includegraphics[width =\figuresize]{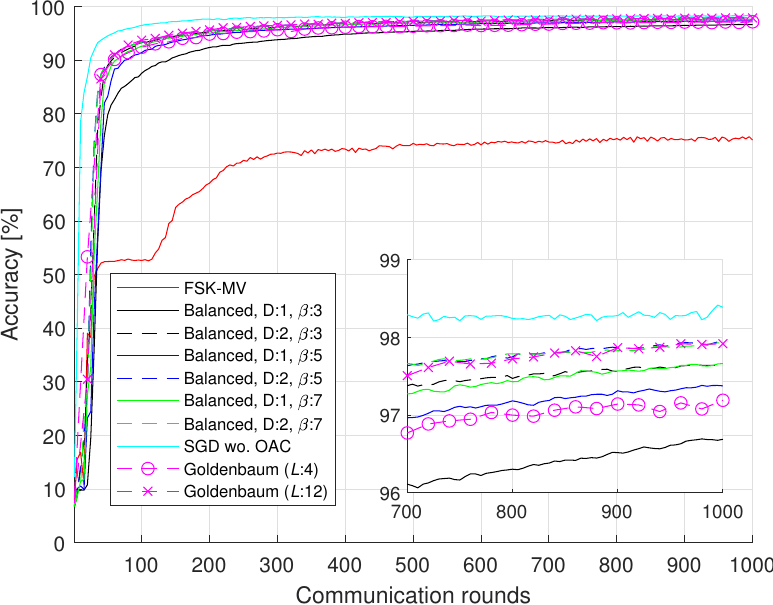}
		\label{subfig:acc_heto_R1_wM_wAAM}}
	\caption{ Test accuracy versus communication rounds (Heterogeneous data distribution, $\numberOfAntennasAtES=1$, $\numberOfEdgeDevices=25$). The proposed scheme with AAM can provide a high test accuracy for a scenario with heterogeneous data distribution.}
	\label{fig:testAccHete}
\end{figure*}
\begin{figure*}
	\centering
	\subfloat[Without AAM (Momentum: 0,  $\valueMaximum=1$).]{\includegraphics[width =\figuresize]{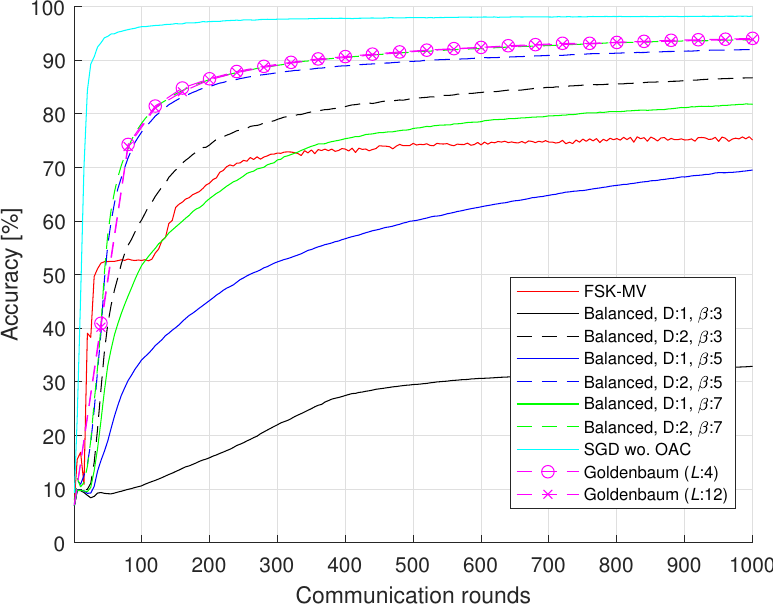}
		\label{subfig:acc_heto_R25_woM_woAAM}}~~~	
	\subfloat[Without AAM (Momentum: 0.9,  $\valueMaximum=1$).]{\includegraphics[width =\figuresize]{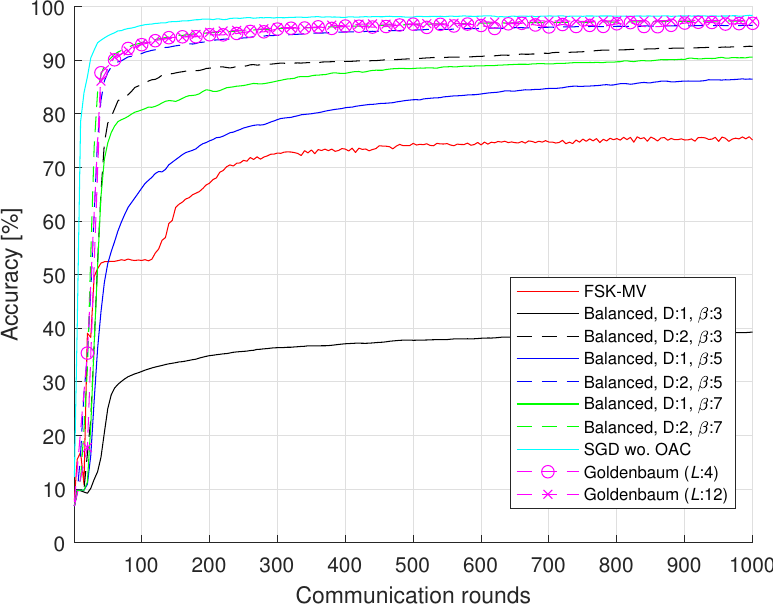}
		\label{subfig:acc_heto_R1_wM_woAAM}}\\		
	\subfloat[With AAM (Momentum: 0).]{\includegraphics[width =\figuresize]{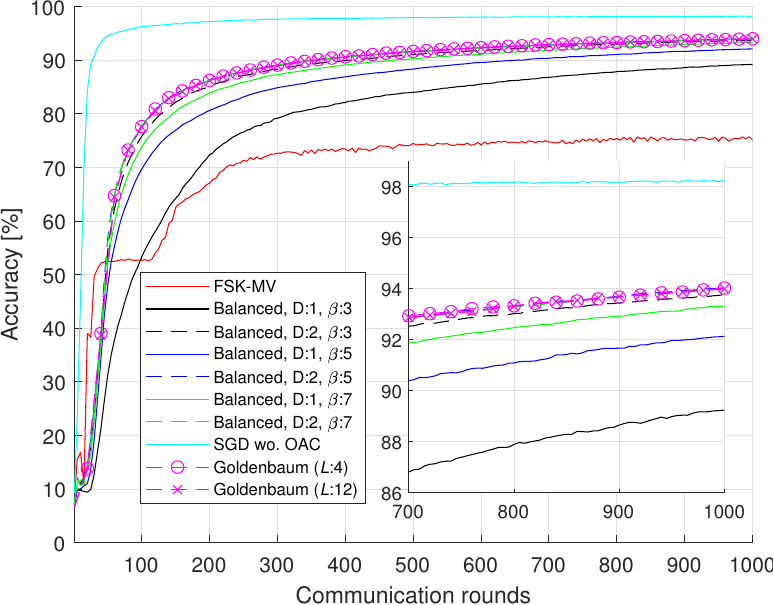}
		\label{subfig:acc_heto_R25_woM_wAAM}}~~~	
	\subfloat[With AAM (Momentum: 0.9).]{\includegraphics[width =\figuresize]{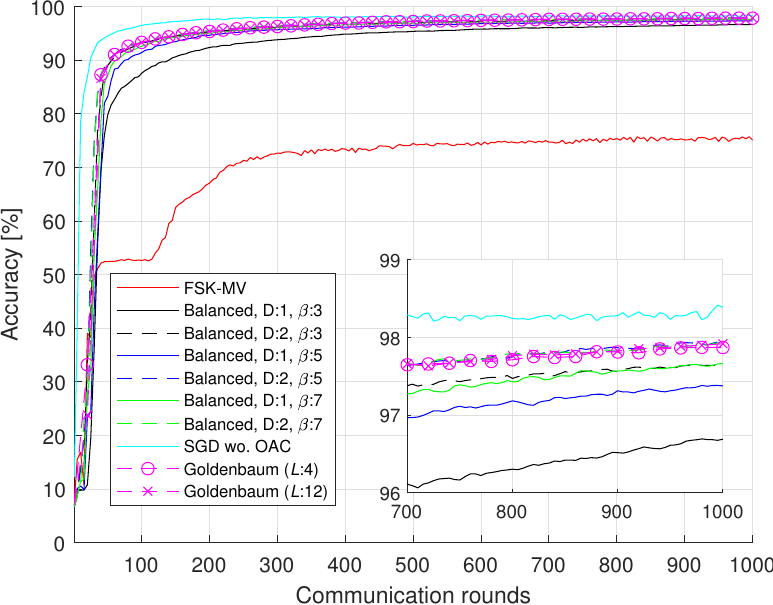}
		\label{subfig:acc_heto_R25_wM_wAAM}}		
	\caption{Test accuracy versus communication rounds (Heterogeneous data distribution, $\numberOfAntennasAtES=25$, $\numberOfEdgeDevices=25$).}
	\label{fig:testAccHeteMoreAntennas}
\end{figure*}
To numerically analyze \ac{OAC} with the proposed scheme for \ac{FEEL}, we consider the learning task of handwritten-digit recognition.
For the fading channel, we consider ITU Extended Pedestrian A (EPA) with no mobility and regenerate the channels between the \ac{ES} and the \acp{ED} independently for each communication round to capture the long-term channel variations.
The subcarrier spacing is set to $15$~kHz. We use  $\numberOfActiveSubcarriers=1200$ subcarriers (i.e., the signal bandwidth is $18$~MHz). Hence, the difference between the time of arrival of the \ac{ED} signals is maximum $\syncError=55.6$~ns. We assume that the synchronization uncertainty at the \ac{ES} is $\Nerror=3$ samples.

For the local data at the \acp{ED}, we use the MNIST database that contains labeled handwritten-digit images size of $28\times28$ from digit 0 to digit 9. We distribute the data samples in the MNIST database to the \acp{ED} to generate representative results for \ac{FEEL}. We consider both homogeneous and heterogeneous data distributions in the cell. To prepare the data, we first choose $|\completeData|=25000$ training images from the database, where each digit has distinct $2500$ images.  
For the scenario with the homogeneous data distribution, we assume that each \ac{ED} has $250$ distinct images for each digit. As done in \cite{sahinCommnet_2021}, for the scenario with  the heterogeneous data distribution, we divide the cell into 5 areas with concentric circles and the \acp{ED} located in $\indexArea$th area have the data samples with the labels $\{\indexArea-1,\indexArea,1+\indexArea,2+\indexArea,3+\indexArea,4+\indexArea\}$ for $\indexArea\in\{1,\mydots,5\}$ (See \cite[Figure 3]{sahinCommnet_2021} for an illustration). The number of \acp{ED} in each area is $5$. As discussed in Section~\ref{sec:system}, we assume that the path loss is compensated through a power control mechanism. For the model, we consider a \ac{CNN}  given  in \cite[Table~I]{Sahin_2022MVjournal}. At the input layer, standard normalization is applied to the data. Our model has $\numberOfModelParameters=123090$ learnable parameters.  For the update rule, the learning rate is set to $0.001$. The batch size $\batchSize$ is set to $64$. To demonstrate the compatibility of the proposed scheme to \ac{SGD} with momentum, we also provide the test accuracy results when the momentum is $0.9$. For the test accuracy calculations, we use $10000$ test samples available in the MNIST database.

In \figurename~\ref{fig:testAccHomo}, we provide the test accuracy versus communication rounds for the scenario with homogeneous data distribution for $\numberOfAntennasAtES=1$ antenna. In \figurename~\ref{fig:testAccHomo}\subref{subfig:acc_homo_R1_woM_woAAM}, the momentum is zero and we do not consider the \ac{AAM} and set $\valueMaximum=1$. For this scenario,  the accuracy results improve  with the proposed scheme  for larger $\base$ or $\numberOfDigits$ (i.e., less $\coefficientOne$). The proposed scheme  without AAM becomes more blind to the gradients over the communication rounds as their magnitudes tend to reduce. The \ac{FSK-MV} is superior to the proposed scheme because \ac{FSK-MV} is based on \ac{signSGD}, while the proposed scheme implements \ac{SGD} and the proposed scheme increases the noise on the gradient estimates as predicted by Theorem~1. In \cite{Bernstein_2018}, it was also mentioned that \ac{signSGD} can outperform \ac{SGD} by providing stronger weight to the gradient direction as compared to \ac{SGD} when the gradients are noisy. Goldenbaum's scheme performs similarly to the proposed scheme for large $\base$ and $\numberOfDigits$. However, since it is based on analog modulation, it is much more robust to quantization errors as compared to the proposed scheme without AAM. In \figurename~\ref{fig:testAccHomo}\subref{subfig:acc_homo_R1_woM_wAAM}, we re-run the simulation with \ac{AAM}. In this case, the convergence rate improves considerably for all $\base$ and $\numberOfDigits$ since \ac{AAM} eliminates the additive noise term due to the proposed scheme in Theorem~1. The performance with the choice of $\{\base=3,\numberOfDigits=1\}$ is worse than the other configurations since the quantization error is dominant in the case. The best performance is obtained with the \ac{FSK-MV} due to its inherent benefits of \ac{signSGD}.
In \figurename~\ref{fig:testAccHomo}\subref{subfig:acc_homo_R1_wM_w0AAM}, \ac{SGD} is used with the momentum. A non-zero momentum improves the convergence rates for all configurations. However, it causes an unstable behavior for Goldenbaum's scheme, which may be due to the skewed error distribution shown in \figurename~\ref{fig:histogram}.
 In \figurename~\ref{fig:testAccHomo}\subref{subfig:acc_homo_R1_wM_wAAM}, we re-evaluate the same configurations with the \ac{AAM}. In this case, both test accuracy and the convergence rate are improved for the proposed scheme. Also, the final test accuracy reaches almost 98\%, better than the one with \ac{FSK-MV}. Similarly, \ac{AAM} also improves Goldenbaum's scheme while addressing the instability.
 
 In \figurename~\ref{fig:testAccHomoMoreAntennas}, we consider $\numberOfAntennasAtES=25$ antennas. Although using more antennas can improve the \ac{BMSE} considerably, its impact on the test accuracy for the proposed scheme is almost negligible. This is because using more antennas reduces the channel noise as in \figurename~\ref{fig:MSE}, but it cannot reduce the quantization noise which is a function of $\valueMaximum$ and the gradient distribution  changing over communication rounds. The results in \figurename~\ref{fig:testAccHomoMoreAntennas} indicate that the proposed scheme can achieve notable test accuracy results if the quantization error is reduced at the expense of more resource consumption even when there is only a {\em single} antenna at the \ac{ES}. As can be seen from \figurename~\ref{fig:testAccHomoMoreAntennas}\subref{subfig:acc_homo_R25_woM_wAAM}, the instability of the Goldenbaum's scheme  in \figurename~\ref{fig:testAccHomo}\subref{subfig:acc_homo_R1_woM_wAAM} is addressed with more antennas.

In \figurename~\ref{fig:testAccHete}, the test accuracy is evaluated when the data distribution is highly heterogeneous, i.e., each \ac{ED} has only 6 unique digits.  In this case, the performance of the \ac{FSK-MV}  degrades drastically, whereas the performance of the proposed scheme is similar to the one in \figurename~\ref{fig:testAccHomo}. The test accuracy under heterogeneous data distribution is less than 80\% for the \ac{FSK-MV} (this is also reported in \cite{sahinCommnet_2021}). This is because of the bias in the \ac{MV} for the heterogeneous data distribution scenario. For example, the digits 0 and 9 are available at fewer EDs, which makes the MV biased. Hence, the training does not learn these digits well.
On the other hand, the proposed scheme with large $\base$ and $\numberOfDigits$ can achieve more than 90\% test accuracy as shown in \figurename~\ref{fig:testAccHete} for $\numberOfAntennasAtES=1$. A similar observation can be made for $\numberOfAntennasAtES=25$ as in \figurename~\ref{fig:testAccHeteMoreAntennas}\subref{subfig:acc_heto_R25_woM_woAAM}-\subref{subfig:acc_heto_R25_wM_wAAM}, i.e., the proposed scheme can provide test accuracy up to 98\% even if the data distribution is not homogeneous.

\begin{figure}
\centering
\subfloat[Homogeneous data distribution.]{\includegraphics[width =\figuresize]{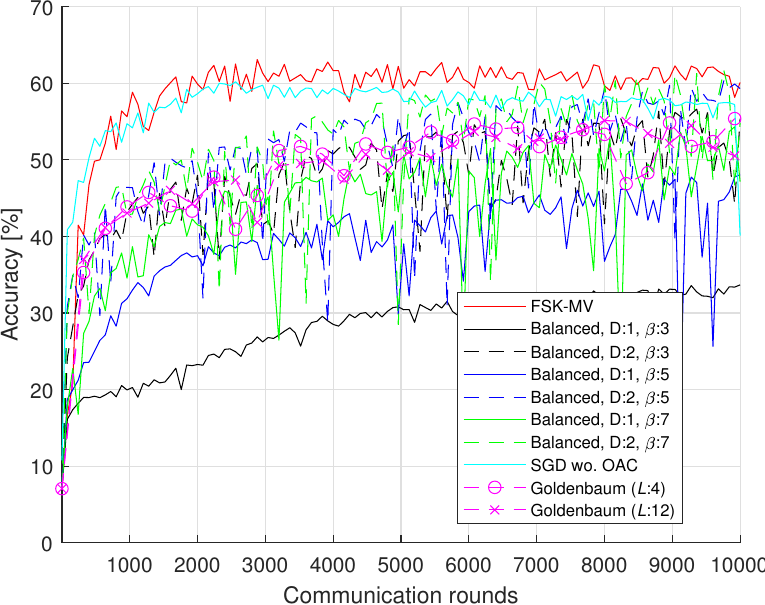}
	\label{subfig:cifarhom}}\\
\subfloat[Heterogeneous data distribution.]{\includegraphics[width =\figuresize]{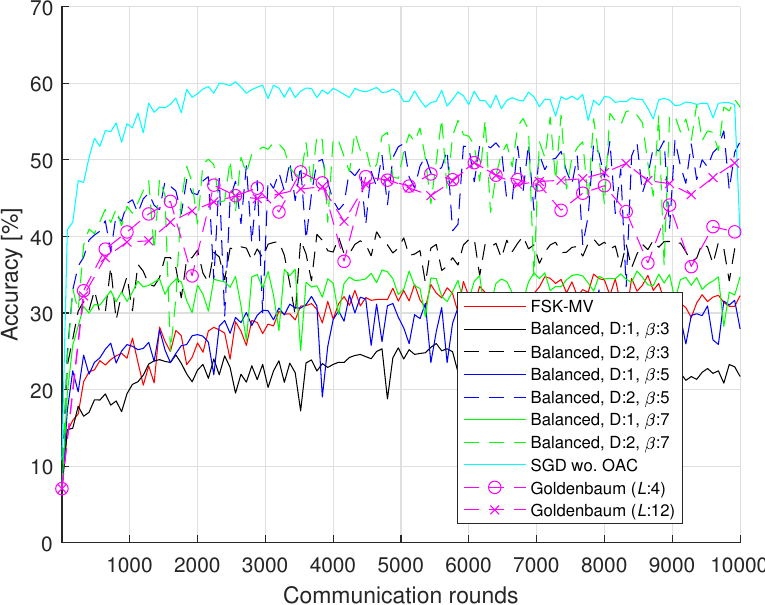}
	\label{subfig:cifarhet}}
\caption{Test accuracy versus communication rounds for CIFAR10 dataset ($\numberOfAntennasAtES=1$, momentum: $0$, AAM is enabled).}
\label{fig:cifar}
\end{figure}	
In \figurename~\ref{fig:cifar}, we provide the simulation results for
CIFAR10 ($|\completeData|=50000$). For CIFAR10, the neural network in \cite[Table~I]{Sahin_2022MVjournal} is extended to three channels, where the size of each channel is 32 by 32. It is well-known that CIFAR10 is a more challenging dataset than MNIST and the test accuracy is less than 70 percent. However, the relative orders of the curves for the proposed scheme  are consistent with ones in \figurename\ref{fig:testAccHomo}\subref{subfig:acc_homo_R1_woM_wAAM} and \figurename\ref{fig:testAccHete}\subref{subfig:acc_heto_R1_woM_wAAM})

\section{Concluding Remarks}
\label{sec:conclusion}

In this study, we investigate an \ac{OAC} method that exploits balanced number systems for gradient aggregation. The proposed scheme achieves a continuous-valued computation through a digital scheme by exploiting the fact that the average of the numerals in the real domain can be used to compute the average of the corresponding real-valued parameters approximately. With the proposed \ac{OAC} method, the local stochastic gradients are encoded into a sequence where the elements of the sequence determine the activated \ac{OFDM} subcarriers. We also use a non-coherent receiver to eliminate the precise sample-level time synchronization, channel estimation overhead, and power instabilities due to the channel inversion techniques. To improve its MSE performance, we also introduce \ac{AAM}. We theoretically analyze its MSE performance and its convergence rate for \ac{FEEL} by considering  both homogeneous and heterogeneous distributions. Our numerical results demonstrate that the test accuracy of the \ac{FEEL} with the proposed scheme using \ac{AAM} can reach up to 98\% even when the \acp{ED} do not have all labels in their data sets. 

The proposed scheme provides a potentially rich area to be investigated. For example, in this study, we consider gradient aggregation. On the other hand, one open question is whether  the proposed scheme can also be utilized for parameter aggregation. Based on our numerical tests, the performance (e.g., test accuracy) can be poor as the neural network may not be tolerant to the errors in the model parameters due to the proposed scheme. Hence, evaluating (and enhancing) the proposed scheme with a noise-tolerant neural network (e.g., quantized neural networks) along with various datasets is an interesting future research direction that can be pursued. 

\appendices
{\section{Derivation of \eqref{eq:varChannel}}
\label{app:derChannel}
The parameter $\coefficientOne$ can be derived as
\begin{align}
	&\coefficientOne\triangleq\expectationOperator[{\left(\meanGradientEleEstimate[\indexCommunicationRound][\indexGradient]-\meanGradientEleOverQuantized[\indexCommunicationRound][\indexGradient]\right)^2}][{\meanGradientEleOverQuantized[\indexCommunicationRound][\indexGradient]}]\nonumber\\
	&\stackrel{(a)}{=}\expectationOperator[\frac{\valueMaximum^2}{\normalizationDigit^2\numberOfAntennasAtES\numberOfEdgeDevices^2}\sum_{\indexDigit=0}^{\numberOfDigits-1}\sum_{\indexActiveSymbol=0}^{\base-2}\symbol[\indexActiveSymbol]^2\left(\numberOFEDsForOptionGeneral[\indexActiveSymbol]+\frac{\noiseVariance}{\symbolEnergy}\right)^2\base^{2\indexDigit}][{\meanGradientEleOverQuantized[\indexCommunicationRound][\indexGradient]}] \nonumber\\
	&\stackrel{(b)}{=}\frac{\valueMaximum^2}{\normalizationDigit^2\numberOfAntennasAtES\numberOfEdgeDevices^2}\sum_{\indexDigit=0}^{\numberOfDigits-1}\sum_{\indexActiveSymbol=0}^{\base-2}\symbol[\indexActiveSymbol]^2\expectationOperator[\left(\numberOFEDsForOptionGeneral[\indexActiveSymbol]+\frac{\noiseVariance}{\symbolEnergy}\right)^2][{\numberOFEDsForOptionGeneral[\indexActiveSymbol]}]\base^{2\indexDigit}\nonumber\\
	&\stackrel{(c)}{=}\frac{\valueMaximum^2}{\numberOfAntennasAtES}\left(\frac{1}{3\base}+\frac{1}{\numberOfEdgeDevices}\left(\frac{\base-1}{3\base}+\frac{2\noiseVariance}{3\symbolEnergy}\right)+\frac{\base\sigma_{\rm n}^4}{3\numberOfEdgeDevices^2\symbolEnergy^2}\right)\frac{\base^\numberOfDigits+1}{\base^\numberOfDigits-1}~,\nonumber\\
	&\stackrel{(d)}{=}\valueMaximum^2 		{\frac{1}{3\numberOfAntennasAtES}\left(\frac{1}{\base} \left(1+\frac{\base\sigma_{\rm n}^2}{\numberOfEdgeDevices(\base-1)}\right)^2+\frac{\base}{\numberOfEdgeDevices(\base-1)}\right)\frac{\base^{\numberOfDigits}+1}{\base^\numberOfDigits-1}}~,\nonumber
\end{align}
where (a) is from \eqref{eq:varGivenK}, (b) is because  the distribution of $\numberOFEDsForOptionGeneral[\indexActiveSymbol]$ is $\binomDist[\numberOfEdgeDevices][1/\base]$ when the distribution of $\localGradientElement[\indexED,\indexGradient][\indexCommunicationRound]$ is $\uniformDist[-\valueMaximumPrime][\valueMaximumPrime]$,  (c) is because of the relation given by
\begin{align}
	&\expectationOperator[\left(\numberOFEDsForOptionGeneral[\indexActiveSymbol]+\frac{\noiseVariance}{\symbolEnergy}\right)^2][{\numberOFEDsForOptionGeneral[\indexActiveSymbol]}]\nonumber\\
	&
	=
	\expectationOperator[\numberOFEDsForOptionGeneral[\indexActiveSymbol]^2][{\numberOFEDsForOptionGeneral[\indexActiveSymbol]}] +2\frac{\noiseVariance}{\symbolEnergy}\expectationOperator[\numberOFEDsForOptionGeneral[\indexActiveSymbol]][{\numberOFEDsForOptionGeneral[\indexActiveSymbol]}]+\frac{\sigma_{\rm n}^4}{\symbolEnergy^2}
	\nonumber
	\\
	&
	=
	\frac{\numberOfEdgeDevices^2}{\base^2}+\frac{\numberOfEdgeDevices(\base-1)}{\base^2} +2\frac{\noiseVariance}{\symbolEnergy}\frac{\numberOfEdgeDevices}{\base}+\frac{\sigma_{\rm n}^4}{\symbolEnergy^2}
	\nonumber
	\\
	&	
	=\frac{\numberOfEdgeDevices^2}{\base^2}+\numberOfEdgeDevices\left(\frac{\base-1}{\base^2}+\frac{2}{\base}\frac{\noiseVariance}{\symbolEnergy}\right)+\frac{\sigma_{\rm n}^4}{\symbolEnergy^2}~,
	\label{eq:MSEpre}
\end{align}
and the identities given by 
\begin{align}
	\frac{1}{\normalizationDigit^2}\sum_{\indexDigit=0}^{\numberOfDigits-1}\base^{2\indexDigit}&=\frac{1}{\normalizationDigit^2}\frac{\base^{2\numberOfDigits}-1}{\base^2-1}=\frac{4}{\base^2-1}\frac{\base^\numberOfDigits+1}{\base^\numberOfDigits-1}~,\nonumber\\
	\sum_{\indexActiveSymbol=0}^{\base-2}\symbol[\indexActiveSymbol]^2&=\sum_{\indexActiveSymbol=0}^{\frac{\base-1}{2}}\indexActiveSymbol^2=\frac{(\base-1)\base(\base+1)}{12}~,\nonumber
\end{align}
for $\normalizationDigit\triangleq{(\base^{\numberOfDigits}-1)}/{2}$, and (d) is because of $\symbolEnergy=\base-1$ as discussed in Section~\ref{subsec:tx}.}

\section{Proof of Theorem~\ref{th:convergence}}
\label{app:proofTh1}
\begin{proof}
By Assumption~\ref{assump:smooth}, we utilize Lemma~1 to obtain the following inequality:
	\begin{align}
		\lossFunctionGlobal[{\modelParametersAtIteration[\indexCommunicationRound+1]}]& - \lossFunctionGlobal[{\modelParametersAtIteration[\indexCommunicationRound]}]\le -\learningRate{\globalGradient[\indexCommunicationRound]}^{\rm T}\meanGradientVectorEstimate[\indexCommunicationRound] +\frac{\learningRate^2\LipschitzConstant}{2}\norm{\meanGradientVectorEstimate[\indexCommunicationRound]}_2^2\nonumber~,
	\end{align}
for $
\modelParametersAtIteration[\indexCommunicationRound+1] = \modelParametersAtIteration[\indexCommunicationRound] - \learningRate  \meanGradientVectorEstimate[\indexCommunicationRound]
$. By using Assumptions~\ref{assump:unbiasAverage} and Assumption~\ref{assump:aveQaun}, we obtain
\begin{align}
	\expectationOperator[{\globalGradient[\indexCommunicationRound]}^{\rm T}\meanGradientVectorEstimate[\indexCommunicationRound]][{}]= {\globalGradient[\indexCommunicationRound]}^{\rm T} \left(\expectationOperator[\meanGradientVector[\indexCommunicationRound]+\biasQuan[\indexCommunicationRound]+\channelNoise[\indexCommunicationRound]][{}]\right)=\norm{\globalGradient[\indexCommunicationRound]}_2^2~.
	\nonumber
\end{align}
By using Assumptions~\ref{assump:graDiv}-\ref{assump:mse}, we can also obtain
\begin{align}
	&\expectationOperator[\norm{\meanGradientVectorEstimate[\indexCommunicationRound]}_2^2][{}]=\nonumber\norm*{\expectationOperator[\meanGradientVectorEstimate[\indexCommunicationRound]][]}_2^2+\expectationOperator[\norm*{\meanGradientVectorEstimate[\indexCommunicationRound]-\expectationOperator[\meanGradientVectorEstimate[\indexCommunicationRound]][]}_2^2][{}]
	\\&=\norm{\globalGradient[\indexCommunicationRound]}_2^2+\underbrace{\expectationOperator[\norm{\meanGradientVectorEstimate[\indexCommunicationRound]-\meanGradientVector[\indexCommunicationRound]}_2^2][{}]}_{\le(\coefficientOne+\coefficientTwo)\numberOfModelParameters}+\underbrace{\expectationOperator[\norm{\meanGradientVector[\indexCommunicationRound]-\globalGradient[\indexCommunicationRound]}_2^2][{}]}_{\le\frac{1}{\numberOfEdgeDevices}\sum_{\indexED=0}^{\numberOfEdgeDevices-1}\secondMoment[\indexDigit]}\label{eq:identityB}~.
\end{align} 
Therefore, for a given $\modelParametersAtIteration[\indexCommunicationRound]$, the expected improvement can be expressed as
	\begin{align}
		&\expectationOperator[{	\lossFunctionGlobal[{\modelParametersAtIteration[\indexCommunicationRound+1]}] - \lossFunctionGlobal[{\modelParametersAtIteration[\indexCommunicationRound]}]}][{}]= -\learningRate{\globalGradient[\indexCommunicationRound]}^{\rm T}\expectationOperator[\meanGradientVectorEstimate[\indexCommunicationRound]][] +\frac{\learningRate^2\LipschitzConstant}{2}\expectationOperator[\norm{\meanGradientVectorEstimate[\indexCommunicationRound]}_2^2][]\nonumber~\\
		 &\le -\learningRate\norm{\globalGradient[\indexCommunicationRound]}_2^2+\frac{\learningRate^2\LipschitzConstant}{2}\left(\norm{\globalGradient[\indexCommunicationRound]}_2^2+(\coefficientOne+\coefficientTwo)\numberOfModelParameters+\frac{1}{\numberOfEdgeDevices}\sum_{\indexED=0}^{\numberOfEdgeDevices-1}\secondMoment[\indexED]\right)\nonumber.
	\end{align}
	By using Assumption~\ref{assump:boundedLoss}, we perform a telescoping sum over the iterations and calculate the expectation over the randomness in the trajectory as
	\begin{align}
	\lossFunctionGlobal[{\modelParametersAtIteration[0]}]-\lossFunctionGlobalMinimum&\ge \lossFunctionGlobal[{\modelParametersAtIteration[0]}]-\expectationOperator[{\lossFunctionGlobal[{\modelParametersAtIteration[\communicationRounds]}]}][]\nonumber\\&=\expectationOperator[{\sum_{\indexCommunicationRound=0}^{\communicationRounds-1}\lossFunctionGlobal[{\modelParametersAtIteration[\indexCommunicationRound]}] - \lossFunctionGlobal[{\modelParametersAtIteration[\indexCommunicationRound+1]}]}][]\nonumber\\&\ge\sum_{\indexCommunicationRound=0}^{\communicationRounds-1}\expectationOperator[{\lossFunctionGlobal[{\modelParametersAtIteration[\indexCommunicationRound]}] - \lossFunctionGlobal[{\modelParametersAtIteration[\indexCommunicationRound+1]}]}][]\nonumber\\&\ge		(-\learningRate+\frac{\learningRate^2\LipschitzConstant}{2})\expectationOperator[{\sum_{\indexCommunicationRound=0}^{\communicationRounds-1}\norm{\globalGradient[\indexCommunicationRound]}_2^2}][]\nonumber\\
	&~~~~+\frac{\learningRate^2\LipschitzConstant\communicationRounds}{2}\left((\coefficientOne+\coefficientTwo)\numberOfModelParameters+\frac{1}{\numberOfEdgeDevices}\sum_{\indexED=0}^{\numberOfEdgeDevices-1}\secondMoment[\indexED]\right)~. \nonumber
\end{align}
	By rearranging the terms, \eqref{eq:convergence}  is reached.
\end{proof}
\section{Proof of Theorem~\ref{th:convergenceAAM}}
\label{app:proofTh2}
\begin{proof}
	The proof of Theorem~\ref{th:convergenceAAM} is similar to that of Theorem~\ref{th:convergence}. We re-evaluate $\expectationOperator[\norm{\meanGradientVectorEstimate[\indexCommunicationRound]-\meanGradientVector[\indexCommunicationRound]}_2^2][{}]$ in \eqref{eq:identityB} under AAM. To this end, let $\bias[\indexCommunicationRound][\indexED]\triangleq\expectationOperator[{\localGradient[\indexED][\indexCommunicationRound]-\globalGradient[\indexCommunicationRound]}][]$ be the bias vector due to data heterogeneity. Based on Assumption~\ref{assump:graDiv},
	\begin{align}
	\expectationOperator[\norm{\localGradient[\indexED][\indexCommunicationRound]}_2^2][]&=\expectationOperator[\norm{\localGradient[\indexED][\indexCommunicationRound]-\globalGradient[\indexCommunicationRound]}_2^2][]-\norm{\globalGradient[\indexCommunicationRound]}_2^2-2{\globalGradient[\indexCommunicationRound]}^{\rm T}\bias[\indexCommunicationRound][\indexED]\nonumber\\
	&\le \secondMoment[\indexED] + \norm{\globalGradient[\indexCommunicationRound]}_2^2 +2{\globalGradient[\indexCommunicationRound]}^{\rm T}\bias[\indexCommunicationRound][\indexED]~.
	\label{eq:bias}
\end{align}
Therefore, based on \eqref{eq:aam}, \eqref{eq:bias}, and by Assumption~\ref{assump:unbiasAverage},
	\begin{align}
	&\expectationOperator[\norm{\meanGradientVectorEstimate[\indexCommunicationRound]-\meanGradientVector[\indexCommunicationRound]}_2^2][{\meanGradientVectorEstimate[\indexCommunicationRound]}]=\configurationFactorTotal\expectationOperator[{\valueMaximum^{(\indexCommunicationRound)}}^2][{\{\localGradient[\indexED][\indexCommunicationRound-1]\}}]\nonumber\\
	&=\metricFactor^2\configurationFactorTotal\expectationOperator[\norm{\metricVector[\indexCommunicationRound-1]}_\infty^2][{\{\localGradient[\indexED][\indexCommunicationRound-1]\}}]\nonumber\\
	&\le\metricFactor^2\configurationFactorTotal\expectationOperator[\norm{\metricVector[\indexCommunicationRound-1]}_2^2][{\{\localGradient[\indexED][\indexCommunicationRound-1]\}}]\nonumber\\
	&=\metricFactor^2\configurationFactorTotal\nonumber\expectationOperator[\sum_{\indexED=0}^{\numberOfEdgeDevices-1}\norm{\localGradient[\indexED][\indexCommunicationRound-1]}_2^2][{\{\localGradient[\indexED][\indexCommunicationRound-1]\}}]\\
	&=\metricFactor^2\configurationFactorTotal\nonumber\sum_{\indexED=0}^{\numberOfEdgeDevices-1}\expectationOperator[\norm{\localGradient[\indexED][\indexCommunicationRound-1]}_2^2][{\localGradient[\indexED][\indexCommunicationRound-1]}]\\
	&\le\metricFactor^2\configurationFactorTotal\left(\numberOfEdgeDevices\norm{\globalGradient[\indexCommunicationRound-1]}_2^2+\sum_{\indexED=0}^{\numberOfEdgeDevices-1}\secondMoment[\indexED]+2{\globalGradient[\indexCommunicationRound-1]}^{\rm T}\underbrace{\sum_{\indexED=0}^{\numberOfEdgeDevices-1}\bias[\indexCommunicationRound-1][\indexED]}_{=0}\right). \nonumber
\end{align}
Therefore,  the expected improvement given in Appendix~\ref{app:proofTh1} with \ac{AAM} can be re-expressed as
	\begin{align}
	&\expectationOperator[{	\lossFunctionGlobal[{\modelParametersAtIteration[\indexCommunicationRound+1]}] - \lossFunctionGlobal[{\modelParametersAtIteration[\indexCommunicationRound]}]}][{}]= -\learningRate{\globalGradient[\indexCommunicationRound]}^{\rm T}\expectationOperator[\meanGradientVectorEstimate[\indexCommunicationRound]][] +\frac{\learningRate^2\LipschitzConstant}{2}\expectationOperator[\norm{\meanGradientVectorEstimate[\indexCommunicationRound]}_2^2][]\nonumber~\\
	&
	~~~~~~~~~~~\le  (-\learningRate+\frac{\learningRate^2}{2}\LipschitzConstant)\norm{\globalGradient[\indexCommunicationRound]}_2^2\nonumber+\frac{\learningRate^2}{2}\LipschitzConstant\metricFactor^2\configurationFactorTotal\numberOfEdgeDevices\norm{\globalGradient[\indexCommunicationRound-1]}_2^2\\&~~~~~~~~~~~~~+\frac{\learningRate^2}{2}\LipschitzConstant(\metricFactor^2\configurationFactorTotal\numberOfEdgeDevices+1)\frac{1}{\numberOfEdgeDevices}\sum_{\indexED=0}^{\numberOfEdgeDevices-1}\secondMoment[\indexED]\nonumber~.
\end{align}
	Considering Assumption~\ref{assump:boundedLoss}, we perform a telescoping sum over the iterations and calculate the expectation over the randomness in the trajectory as 
	\begin{align}
	\lossFunctionGlobal[{\modelParametersAtIteration[1]}]-\lossFunctionGlobalMinimum&\ge\sum_{\indexCommunicationRound=0}^{\communicationRounds-1}\expectationOperator[{\lossFunctionGlobal[{\modelParametersAtIteration[\indexCommunicationRound]}] - \lossFunctionGlobal[{\modelParametersAtIteration[\indexCommunicationRound+1]}]}][]\nonumber\\
	&\ge
	(-\learningRate+\frac{\learningRate^2\LipschitzConstant}{2})\expectationOperator[{\sum_{\indexCommunicationRound=1}^{\communicationRounds}\norm{\globalGradient[\indexCommunicationRound]}_2^2}][]\nonumber\\
	&~~~+\frac{\learningRate^2}{2}\LipschitzConstant\metricFactor^2\configurationFactorTotal\numberOfEdgeDevices\expectationOperator[{\sum_{\indexCommunicationRound=1}^{\communicationRounds}\norm{\globalGradient[\indexCommunicationRound-1]}_2^2}][]\nonumber\\
	&~~~+\frac{\learningRate^2\LipschitzConstant\communicationRounds}{2}(\metricFactor^2\configurationFactorTotal\numberOfEdgeDevices+1)\frac{1}{\numberOfEdgeDevices}\sum_{\indexED=0}^{\numberOfEdgeDevices-1}\secondMoment[\indexED]~. \label{eq:final}
\end{align}
Also, we can express the expected value of the sum over the trajectory as
\begin{align}
\expectationOperator[{\sum_{\indexCommunicationRound=1}^{\communicationRounds}\norm{\globalGradient[\indexCommunicationRound-1]}_2^2}][]=\expectationOperator[{\sum_{\indexCommunicationRound=1}^{\communicationRounds}\norm{\globalGradient[\indexCommunicationRound]}_2^2}][]+\expectationOperator[{\norm{\globalGradient[0]}_2^2-\norm{\globalGradient[T]}_2^2}][]~.\label{eq:sumt}
\end{align}
Finally, by using \eqref{eq:sumt} and rearranging the terms in \eqref{eq:final}, \eqref{eq:convergence}  is obtained.
\end{proof}

\acresetall

\bibliographystyle{IEEEtran}
\bibliography{references}

\end{document}